\pdfoutput=1
\documentclass{aa}  

\usepackage[normalem]{ulem}

\usepackage{graphicx}
\usepackage{txfonts}
\usepackage{slashbox}
\usepackage{mhchem}
\usepackage{xcolor}
\usepackage[normalem]{ulem}

\definecolor{deepmagenta}{rgb}{0.8, 0.0, 0.8}

\def\Halpha{\mbox{H$\alpha$}\xspace}
\def\Hbeta{\mbox{H$\beta$}\xspace}
\def\CaIR{\mbox{\ion{Ca}{II}\,8542~\AA}\xspace}
\def\CaK{\mbox{\ion{Ca}{II}\,K}\xspace}
\def\CaH{\mbox{\ion{Ca}{II}\,H}\xspace}
\def\CaHK{\mbox{\ion{Ca}{II}\,H\,\&\,K}\xspace}
\def\K3{\mbox{K$_3$}\xspace} 
\def\K2{\mbox{K$_2$}\xspace}
\def\K2v{\mbox{K$_{2V}$}\xspace}
\def\K2r{\mbox{K$_{2R}$}\xspace}
\def\Mghk{\mbox{\ion{Mg}{II}\,h\,\&\,k}\xspace}

\def\NaD1{\mbox{Na\,D$_1$}\xspace}
\def\NaD2{\mbox{Na\,D$_2$}\xspace}
\def\NaD12{\mbox{Na\,D$_1$\,\&\,D$_2$}\xspace}

\usepackage{afterpage}

\usepackage[colorlinks=true,linkcolor=blue,citecolor=blue,urlcolor=blue]{hyperref}

\usepackage{siunitx}

\newcommand{\com}[1]{{\color{orange}{#1}}}

\begin{document} 

\title{A comparative study of two X2.2 and X9.3 solar flares observed with HARPS-N}

\subtitle{Reconciling Sun-as-a-star spectroscopy and high-spatial resolution 
    solar observations in the context of the solar-stellar connection}

\author{%
    A.G.M. Pietrow\inst{1} 
    \and M. Cretignier\inst{2} 
    \and M. K. Druett\inst{3} 
    \and J. D. Alvarado-G\'omez\inst{1} 
    \and S. J. Hofmeister\inst{1}
    \and M. Verma\inst{1}
    \and R. Kamlah\inst{1,4}
    \and M. Baratella\inst{1,5}
    \and E. M. Amazo-G\'omez\inst{1}
    \and I. Kontogiannis\inst{1}
    \and E. Dineva\inst{3,1}
    \and A. Warmuth\inst{1}
    \and C. Denker\inst{1}
    \and K. Poppenhaeger\inst{1,4}
    \and O.~Andriienko\inst{6}
    \and X. Dumusque\inst{7}
    \and M. G. L{\"o}fdahl\inst{6}}

\institute{%
    \inst{1}Leibniz-Institut für Astrophysik Potsdam (AIP), An der Sternwarte 16, 14482 Potsdam, Germany\\   
    \inst{2}Department of Physics, University of Oxford, OX13RH Oxford, United Kingdom \\
    \inst{3}Centre for mathematical Plasma Astrophysics, KU Leuven, Celestijnenlaan 200B, 3001 Leuven, Belgium\\
    \inst{4}Universit{\"a}t Potsdam, Institut f{\"u}r Physik und Astronomie, Karl-Liebknecht-Stra{\ss}e 24/25, 14476~Potsdam, Germany\\
    \inst{5}European Southern Observatory, Alonso de Cordova 3107 Vitacura, Santiago de Chile, Chile\\
    \inst{6}Institute for Solar Physics, Dept. of Astronomy, Stockholm University, Albanova University Centre, SE-106 91 Stockholm, Sweden\\
    \inst{7}Astronomy Department, University of Geneva, 51 ch. de Pegasi, 1290 Versoix, Switzerland\\
    \email{apietrow@aip.de}}

\date{Received September 7, 2023; accepted November 14, 2023}

\abstract%
    {Stellar flares cannot be spatially resolved, which complicates ascertaining the physical processes behind particular spectral signatures. Due to their proximity to Earth, solar flares can serve as a stepping stone for understanding their stellar counterparts, especially when using a Sun-as-a-star instrument and in combination with spatially resolved observations.}
    {We aim to understand the disk-integrated spectral behaviors of a confined X2.2 flare and its eruptive X9.3 successor, which had energies of $2.2 \times 10^{31}$~erg and $9.3 \times 10^{31}$~erg, respectively, as measured by Sun-as-a-star observations with the High Accuracy Radial velocity Planet Searcher for the Northern hemisphere (HARPS-N).}
    {The behavior of multiple photospheric (\NaD12, \ion{Mg}{I} at 5173~\AA, \ion{Fe}{I} at 6173~\AA, and \ion{Mn}{I} at 4031~\AA) and chromospheric (\CaHK, \Halpha, \Hbeta, and \ion{He}{I}\,D$_3$) spectral lines were investigated by means of activity indices and contrast profiles. A number of different photospheric lines were also investigated by means of equivalent widths, and radial velocity measures, which were then related to physical processes directly observed in high-resolution observations made with the Swedish 1-meter Solar Telescope (SST) and the Atmospheric Imaging Assembly (AIA) on board of the Solar Dynamics Observatory (SDO).}
    {Our findings suggest a relationship between the evolving shapes of contrast profile time and the flare locations, which assists in constraining flare locations in disk-integrated observations. In addition, an upward bias was found in flare statistics based on activity indices derived from the \CaHK lines. In this case, much smaller flares cause a similar increase in the activity index as that produced by larger flares. \Halpha-based activity indices do not show this bias and are therefore less susceptible to activity jitter. Sodium line profiles show a strongly asymmetric response during flare activity, which is best captured with a newly defined asymmetrical sodium activity index. A strong flare response was detected in \ion{Mn}{I} line profiles, which is unexpected and calls for further exploration. Intensity increases in \Halpha, \Hbeta, and certain spectral windows of AIA before the flare onset suggest their potential use as short-term flare predictors.}
    {}

{}


\maketitle

\section{Introduction}
As we enter the era of detailed studies of exoplanet atmospheric characterization and habitability, there is a growing interest in understanding the environment in which planets are formed and evolve. 
In the case of our Solar System, it is generally accepted that transient energetic phenomena, such as flares, coronal mass ejections (CMEs), and solar energetic particle (SEP) events, are the most important drivers of space weather. Based on this paradigm, this is likely also the case in other systems, especially in those with more active stars. In fact, stellar counterparts of such transient events have the potential to permanently alter the atmospheric conditions of an exoplanet \citep[e.g.,][]{Pulkkinen2007,Temmer2021}. Thus, such events are vitally important in understanding the habitable zone of exoplanets \citep{Airapetian2020}. 

In the stellar regime, survey missions such as the Kepler space telescope \citep{2010Sci...327..977B} and the Transiting Exoplanet Survey Satellite \citep[TESS,][]{2015JATIS...1a4003R}, as well as dedicated observations from the Hubble Space Telescope \citep[HST,][]{1986SPIE..627..350W}, the XMM-Newton mission \citep{1985ntvh.book..291K}, and the Chandra X-ray telescope \citep{2000ChNew...7...11W}, have enabled a relatively robust multi wavelength characterization of the flaring behavior of stars across spectral types and ages \citep[e.g.,][]{Guarcello2019, Davenport2019, MacGregor2021}. Particularly at early stages of stellar evolution and in later spectral types, flare rates, and energies have been observed, exceeding the solar levels by several orders of magnitude \citep[e.g.,][]{Loyd2018a, Loyd2018b, Ilin2019, Ilin2021, Pietras2022}. On the other hand, knowledge of the eruptive behavior of stars (for example, CMEs) and their high-energy particle environment is nowadays extremely limited \citep[e.g.,][]{Leitzinger2022}. The large majority of studies are currently based on direct extrapolations from solar data \citep[e.g.,][]{Kay2016, Kay2019, Patsourakos2017, Hazra2022}, whose validity is not well established in the stellar regime. One of the reasons is that the expected elevated CME activity on active flare stars is not only at odds with multiple reports of non-detections by dedicated multi wavelength observing programs \citep[e.g.,][]{Crosley2016, Villadsen2017, Crosley2018, Vida2019, Muheki2020a, Muheki2020b, Koller2021} but also with the current knowledge on the mass loss budget of cool stars \citep{Drake2013, Odert2017, Wood2021}. In addition, the scarce set of currently known stellar CME-candidate events displays a large deficit of kinetic energy, deviating significantly from the behavior observed in solar flare--CME events \citep[e.g.,][]{Moschou2017, Moschou2019, Vida2019, Leitzinger2022}. The same behavior is observed in the handful of stellar CME events detected so far \citep{Houdebine1990,Gunn1994,Houdebine1993,Guenther1997,Fuhrmeister2004,Vida2016,Argiroffi2019, Namekata2021, Veronig2021}. 

Even less is known about the energetic particle environment of active stars, given the difficulties involved in their detection by remote observations \citep[e.g.,][]{Drake2019}. Regarding the flare emission itself, there are strong divergences when solar versus stellar phenomena are explored and compared in photometric studies in the visible spectrum (for example, white-light flare studies). Flares are commonly detected on Sun-like stars, for example in the TESS and Kepler band-passes \citep[e.g.,][]{2012Natur.485..478M, 2021MNRAS.505L..79Y}, while the detection of such energetic events is very scarce and difficult on the Sun when observed in disk-integrated broadband images \citep[e.g.,][]{2004AAS...204.0215K, 2006JGRA..11110S14W, 2010NatPh...6..690K}. The main reasons for these discrepancies are presumably the age and activity stage of the Sun as compared to younger and more active stars \citep[e.g.,][]{2021ApJ...906...72O}. 

Equivalent width studies of stellar spectra have now reached a point, where modern instrumentation has become sensitive enough to register the changes induced by activity on the average disk-integrated shape of some spectral lines, in particular those forming in the upper layers of the photosphere and above \citep{2019galarza, 2020baratella_gaps, 2020baratella_ges, 2020spina}. This has important implications for the derivation of the stellar atmospheric parameters, which vary over the course of an activity cycle. Despite large efforts, the question remains as to whether these effects are due to dark or bright spots, flares, plages, and magnetic fields or a combination of all \citep[e.g.][]{Cretignier2023}. Thus, studying the variability of the equivalent width and other spectral parameters during such events is of primary importance.



Theoretical and numerical investigations study properties of stellar CMEs and energetic particles, as well as different physical conditions that could explain deviations from their solar counterparts \citep[e.g.,][]{Alvarado-Gomez2022}. These deviations can be due to large-scale magnetic fields, which may confine CMEs \citep{Alvarado-Gomez2018, Alvarado-Gomez2019, Alvarado-Gomez2020} and energetic particles \citep{Fraschetti2019, Fraschetti2022}, or prevent triggers of an eruption in the first place \citep{Sun2022}. Progress in this area is however hampered by the lack of observations, especially in the stellar regime, and by large degeneracies and uncertainties \citep[e.g.,][]{Lynch2022}.

One possibility to overcome the aforementioned limitations lies in so-called `Sun-as-a-star' studies, where stellar instrumentation and observing techniques are used to characterize the Sun. Such investigations are getting more traction in recent years, including topics related to radial velocity variations \citep[e.g.,][]{Dumusque(2021), AlMoulla(2023)}, stellar activity and magnetic fields \citep[e.g.,][]{Milbourne2019, Maldonado2019, Thompson2020}, convection patterns \citep{Miklos2020}, and total solar irradiance \citep{Milbourne2021}. 


In the context of flares and CMEs, \citet{Namekata2021} and \citet{Otsu_2022} provided a solid starting point for capturing different line-profile variations and their interpretation in connection with the characteristics of the underlying transient event. However, these studies do not completely fulfill the Sun-as-a-star condition, as spatially resolved information is used to extract faint signals from data, and large-scale effects such as other activity signatures and center-to-limb variations (CLV) are not taken into account. Thus, these findings should be compared to  `true' Sun-as-a-star observations to see how they hold up in this regime \citep[e.g. ][]{PietrowNessi}. Nevertheless, the resolved context offered in the prior works acts as a bridge between solar and stellar observations and offers ways to `translate' results back and forth between the fields. 

One important aspect of such a translation is to find a link between solar and stellar notations of flare strengths. In solar physics, the strength of solar flares is typically expressed using a logarithmic scale \citep{Baker1970}, which is based on the peak intensity of the X-ray emission in the 1\,--\,8~\AA\ band. Flares are given one of five designations (A, B, C, M, or X), which corresponds to the exponent, and a number between 1 and 9, which acts as a multiplication factor (for example, an M5.0 flare corresponds to $5\times10^{-5}$~W~m$^{-2}$). X-class flares are an exception to this scaling scheme, as they are defined to be the strongest category and thus open-ended in intensity \citep[][p.~28]{Pietrowthesis}. Attempts have been made to link these classes to bolometric energies, where estimates place an X1.0 flare at $10^{31}$~erg \citep[e.g.,][]{Shibata2013, Maehara2015, Namekata2021}. However, such relations do not scale linearly \citep[e.g.,][Fig.~8]{Warmuth2016}. The reason for this is not known and requires further research. Therefore, the linear relation will be used for simplicity in our study. The two flares studied in this work have been measured to be X2.2 and X9.3 respectively, which corresponds to a rough energy of $2.2 \times 10^{31}$~erg and $9.3 \times 10^{31}$~erg, respectively. This is at the lower end of the $10^{31} - 10^{36}$~erg energy range of flares described by \citet{Pietras2022}, with the average flare being around X100, or 10 times stronger than the large flare described in this work.

This work focuses on Sun-as-a-star observations of two large solar flares observed with the solar telescope \citep{Dumusque(2015)} connected to the High Accuracy Radial-velocity Planet Searcher for the Northern hemisphere \citep[HARPS-N,][]{Cosentino2012} instrument of the Telescopio Nazionale Galileo (TNG) on La Palma. Later extensive solar context based on instruments with a high spatial resolution is added to demonstrate what can and cannot be seen in the disk-integrated HARPS-N spectra, as a way to highlight both the power and limitations of 1-D flare observations.

\section{Observations}\label{sect:obs}
\begin{figure*}
	\centering
	\includegraphics[width=\textwidth]{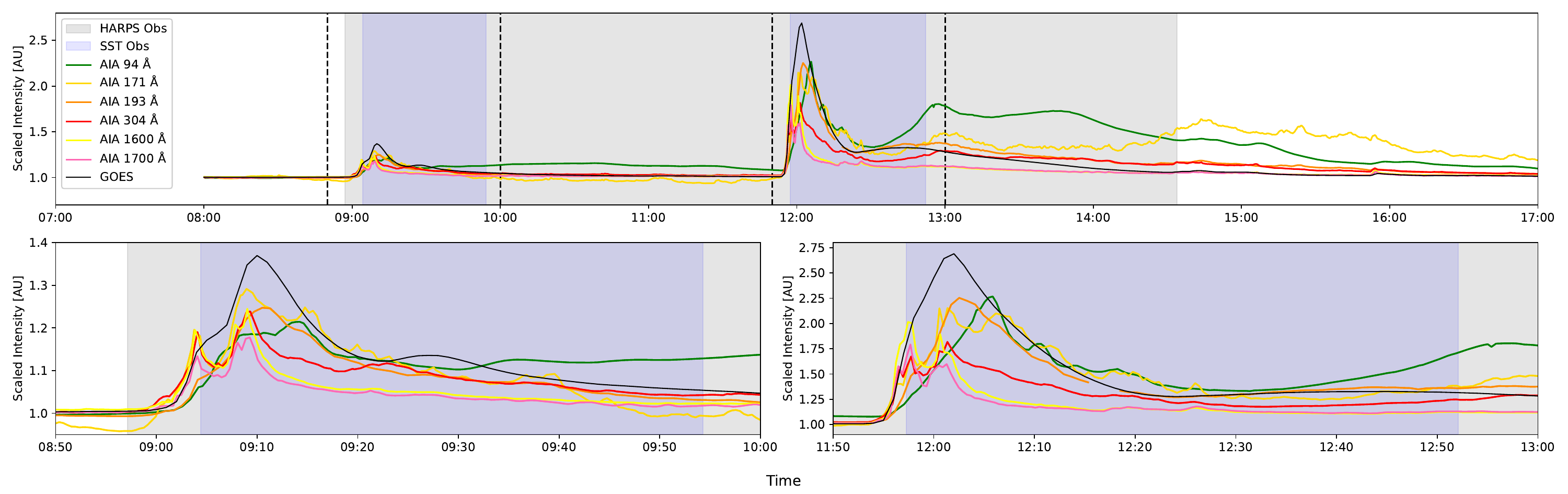}
	\caption{Comparison between spatially integrated and normalized AIA UV and EUV intensities and the normalized GOES X-ray flux. The area between the dashed lines shows the time window over which the X2.2 and X9.3 flares have been enlarged for the lower two panels. The gray shaded area represents the time during which HARPS-N observed, and the blue shaded area represents when SST was observing.}
	\label{Fig:aia}
\end{figure*}


The primary dataset used in this work is the HARPS-N time series, with all other observations being used to give context to these observations. 

\subsection{Sun-as-a-star spectra}

Sun-as-a-star spectra\footnote{\tiny{The raw data can be accessed through this link: \url{https://plone.unige.ch/HARPS-N/harps-n-operations-and-observations/harps-n-solar-telescope-data-release}}} were observed between 08:57 and 14:33~UT on the 6~September 2017 (shaded gray in Fig.~\ref{Fig:aia}), with two flares taking place between 8:57 and 9:17~UT and between 11:53 and 12:10~UT respectively. They were observed with the 3-inch solar telescope \citep{Dumusque(2015)} that feeds into HARPS-N. The observations were made at a 5-minute cadence, recording the visible spectrum between 3900 and 6900~\AA\ with a spectral resolution of \mbox{${\cal R} \approx 110\,000$}. HARPS-N is a cross-dispersed echelle spectrograph, which observes several dozens of orders at once on a \mbox{4k\,$\times$\,4k}-pixel CCD mosaic before they are merged by the standard data reduction software of HARPS-N to produce a one-dimensional spectrum of the full spectral range. 

The HARPS-N spectra are usually not photometrically calibrated since conservation of the \'etendue requires small optical fibers to collect the stellar light at the high-spectral resolution, whereby the sky-projected size of the fibers are similar to the on-site seeing disk \citep{Pepe(2021)}. Furthermore, other airmass effects such as differential extinction or weather conditions changes are not corrected. Thus, the shape of the spectrum is not preserved due to the different filters inside the instruments that try to balance the overall flux in order to avoid saturation of the detector. All of these features, combined with guiding errors, result in flux variations of up to 50\% \citep{Dumusque(2021)}. This means that the data does not contain reliable flux values, and has to be studied after the continuum is fitted. 

In this work, this was achieved with the Rolling Alpha Shape for a Spectrally Improved Normalisation Estimation \citep[RASSINE,][]{Cretignier(2020b)} code, which is publicly available on GitHub.\footnote{\tiny\url{https://github.com/MichaelCretignier/Rassine_public}} The code is based on the convex hull theory to fit upper envelopes to a cloud of data point. This filter aims to separate photospheric lines (high-frequency terms) from the continuum (low-frequency term). In more descriptive terms, the continuum is fitted by rolling a circle along the top of the spectrum with a radius larger than the width of the photospheric lines. Each contact point between the `rolling pin' and the spectrum is used to fit the upper envelope of the spectrum. 

Additionally, the HARPS-N spectra are contaminated by several systematic effects such as interference patterns produced by the filters inside the instrument, ghosts that are internal reflections inside the spectrograph, and telluric lines. We used spectra processed with the YARARA software \citep{Cretignier(2021)}, which was designed to correct the time variation of all these effects. Moreover, YARARA features an algorithm to improve the radial velocity precision, which is mainly affected by stellar activity \citep[e.g.,][]{Cretignier(2022)}. However, we disabled the stellar activity correction for this study.

\begin{figure}[t]
	\centering
	\includegraphics[width=9cm]{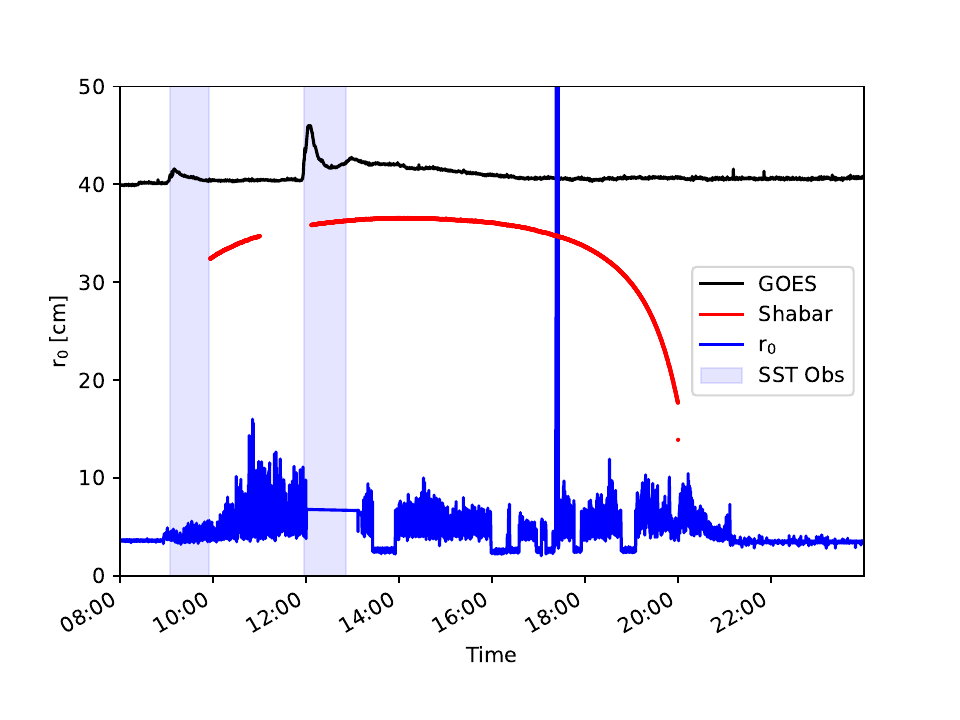}
	\caption{Overview of the weather and seeing conditions at the SST on 6~September 2017. The red curve represents the solar intensity in arbitrary units as measured by the SHABAR showing a smooth curve, which indicates a cloud-free day. The Fried parameter $r_0$ is plotted in blue. The black curve represents the scaled GOES X-ray flux, and the vertical light-blue bands in the background refer to times when the SST flare data were taken.}
	\label{Fig:seeing}
\end{figure}

\subsection{High-resolution observations}
Both flares were also observed by the 1-meter Swedish Solar Telescope \citep[SST,][]{Scharmer03}, using simultaneously the CRisp Imaging SpectroPolarimeter \citep[CRISP,][]{Scharmer08} and the CHROMospheric Imaging Spectrometer \citep[CHROMIS,][]{Scharmer17}. Both instruments are dual Fabry-P\'erot interferometers, which can be tuned to specific wavelengths in order to `scan' through a spectral line with an average spectral resolution of around ${\cal R} \approx 130\,000$. CRISP executed a sequence that alternated between \Halpha and \CaIR, while CHROMIS only recorded the \CaK line together with the 4000~\AA\ continuum.

The cadence was 15~s for CRISP, and the observing sequence consisted of 13 wavelength positions in the \Halpha line at $\pm$1.50, $\pm$1.0, $\pm$0.80, $\pm$0.60, $\pm$0.30, $\pm$0.15, and 0.00~\AA\ relative to line center. The \CaIR scan consisted of 11 wavelength positions taken in full-Stokes-polarimetry mode at $\pm$0.7, $\pm$0.5, $\pm$0.3, $\pm$0.2, $\pm$0.1, and 0.0~\AA\ relative to line center. The CRISP plate scale is 0.058\arcsec~pixel$^{-1}$.

The cadence of the CHROMIS observing sequence was 6.5~s with 19 wavelength positions in the \CaK line at $\pm$1.00,
$\pm$0.85, $\pm$0.65, $\pm$0.55, $\pm$0.45, $\pm$0.35, $\pm$0.25, $\pm$0.15, $\pm$0.07, and 0.00~\AA\ relative to the line center, plus a single continuum point at 4000~\AA. The CHROMIS plate scale is 0.0375\arcsec~pixel$^{-1}$. In addition to the narrow-band images, wide-band images were obtained co-temporally with each CRISP and CHROMIS narrow-band exposure for alignment purposes.

The X2.2 flare was observed between 09:04 and 09:54~UT, and the field-of-view (FOV) was centered at heliocentric coordinates $(x,\, y) = (532\arcsec, -233\arcsec)$ which corresponds to a latitude of 37$^\circ$ ($\mu$ = 0.79). The X9.3 flare was observed between 11:55 and 12:52~UT, and the FOV was centered at heliocentric coordinates $(x,\, y) = (537\arcsec,\, -222\arcsec)$ which also corresponds to a latitude of 37$^\circ$ ($\mu$ = 0.79). Both time series are shown as a blue-shaded region in Fig.~\ref{Fig:aia}.

The data was processed using the standard SSTRED pipeline \citep{jaime15, mats21} using multi-object multi-frame blind deconvolution \citep[MOMFBD,][]{mats02,vanNoort05}. However, the routine does not perform well under mediocre seeing conditions and can even introduce artifacts in frames with bad seeing conditions \citep[e.g. see Fig. 5.4 of ][]{Pietrowthesis}. In frames where image restoration failed, restored data were replaced simply by the sum of the narrow-band images after applying calibrations for pixel bias and gain (i.e., darks and flats), as well as the calibration for camera alignment. During the summing process shift vectors were derived by cross-correlation on the broadband frames and subsequently applied to the simultaneous narrow-band frames \citep{Pietrow23}. Finally, absolute wavelength and intensity calibrations were performed using the solar spectral atlas by \cite{Neckel1984}.

The \CaIR data of the X2.2 flare was already described in \citet{2021Vissers} but have now been re-processed together with the \Halpha and \CaK observations. The \Halpha and \CaIR data of the X9.3 flare were first described in \citet{Quinn19}, while the \CaK data were first shown in \citep{Pietrow22} and were re-processed for this work.

\subsection{Weather and seeing conditions}

Both the SST and HARPS-N data were taken in close proximity of each other on the Canary Island of La Palma. A video recorded by an all-sky camera indicated a cloud-free day.\footnote{\tiny\url{http://www.sst.iac.es/sky/movies/2017/sky_2017-09-06.mp4}} However, the seeing conditions were variable. A summary of the weather and seeing conditions is provided in Fig.~\ref{Fig:seeing}, showing the temporal evolution of the Fried parameter $r_0$ \citep{Fried66}, the solar intensity measured by the SHAdow BAnd Ranger \citep[SHABAR,][]{Sliepen2010}, and the GOES X-ray flux. SHABAR observations were interrupted for about an hour before the second flare. The adaptive optics (AO) system was switched off starting around 12:00~UT during the second flare, resulting in a fixed value of $r_0 \approx 7$~cm. Other intervals of constant $r_0\approx 3$~cm occurred at times when the AO introduced randomized motions for taking flat-field frames. The peak of $r_0 > 50$~cm signifies a calibration step that involves focusing on a pinhole array.

\begin{figure}
    \centering
    \includegraphics[width=\columnwidth]{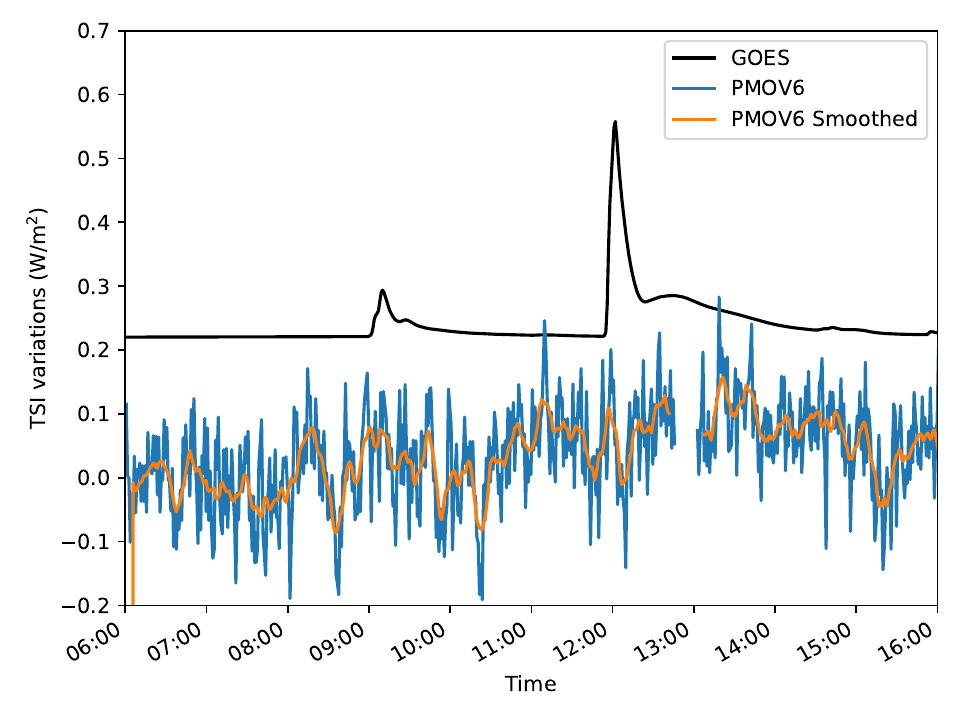}
    \caption{TSI variation during on 6~September 2017 at a 60-second cadence (\textit{blue}) and a corresponding smoothed curve over 10 time steps (\textit{orange}). While both flares are clearly visible in the scaled GOES X-ray flux (\textit{black}), a TSI flare signature is absent. The standard deviation of the PMOV6-A time series is 0.06~W~m$^{-2}$, and the mean TSI is 1359.50~W~m$^{-2}$).}
    \label{Fig:VIRGO-GOES}
\end{figure}
\subsection{Total solar irradiance and X-ray observations}


The Variability of IRradiance and Gravity Oscillation \citep[VIRGO,][]{1997SoPh..170....1F} experiment onboard the Solar and Heliospheric Observatory \citep[SOHO,][]{Domingo95} has been monitoring the total solar irradiance (TSI) since 1996 from the Sun-Earth L$_0$ point \citep[e.g.,][]{pap1999, 2012Metro..49S..34F,Finsterle2021}. The data were averaged resulting in a 1-minute cadence and corrected with the new scale of TSI VIRGO/PMO6V-A \cite[see Discussion, Table 2,][]{Finsterle2021}. The TSI data are available from the VIRGO Team through PMOD/WRC,\footnote{\tiny\url{ftp://ftp.pmodwrc.ch/pub/data/irradiance/virgo/TSI/}} and the SoHO archives at ESA and NASA. The GOES data are provided by NOAA.\footnote{\tiny\url{https://www.ngdc.noaa.gov/stp/satellite/goes-r.html}}

In Fig.~\ref{Fig:VIRGO-GOES}, the TSI is plotted for a large part of the day. The data are plotted with a 60s cadence in blue with a curve that smoothens the data over 10-time steps in orange. A regular jump in the detected signal took place around 13:00~UT and this data has been removed from the plot. 

In addition, a 1\,--\,8~\AA\ Sun-as-a-star integrated light curve taken by the EUV and X-ray Irradiance Sensors \citep[EXIS, ][]{Machol2020} onboard the Geostationary Operational Environmental Satellite No.~16 \citep[GOES-16,][]{2005BAMS...86.1079S, 2017BAMS...98..681S, 2019E&SS....6.1730S} is plotted in black. The X-ray band is sensitive to the solar corona and contains a combination of spectral lines and continuum that are primarily sensitive to flare emission but can also detect quiescent active regions during solar maximum \citep{White2005, Simoes2015}. 

\subsection{Full-disk UV and EUV observations}\label{sec:aia}
The Helioseismic and Magnetic Imager \citep[HMI,][]{2012ScherrerHMI} and the Atmospheric Imaging Assembly \citep[AIA][]{2012LemenAIA} onboard the Solar Dynamics Observatory \citep[SDO, ][]{Pesnell2012} observe the full solar disk in the visible continuum, ultraviolet (UV), and extreme-ultraviolet (EUV) at a cadence of 45~s, 24~s, and 12~s, respectively, and a plate scale of 0.5\arcsec~pixel$^{-1}$ and 0.6\arcsec~pixel$^{-1}$, respectively. HMI is an imaging spectropolarimeter, which records the \ion{Fe}{I} line at 6173~\AA\ in each polarimetric state at five wavelength points and additionally records one point in the continuum \citep{Svanda2018}. The AIA instrument consists of four telescopes that observe in total one continuum, two UV, and seven EUV windows. Each of the AIA channels has a width of several {\AA}ngstr\"oms and thus observes the combined emission of several spectral lines. Consequently, each of the AIA channels is sensitive to a wide range of temperatures, and the temperature response function of each of the channels has multiple peaks related to different atmospheric conditions. The UV channels cover spectral lines from the photosphere and chromosphere \citep{Sim_es_2019}. The EUV channels cover spectral lines in the transition region and solar corona \citep[e.g.,][]{Tr_bert_2014}. During flares, AIA images with standard exposure times typically saturate and AIA switches to a short-exposure mode, where AIA predicts reasonably short exposure times for every other image. Ideally, in the short exposure images, the flaring region does not saturate anymore, but at the same time, the photon counts in the non-flaring (that is, quiescent) regions diminish and are often even below the digitization threshold. Furthermore, AIA observations suffer from strong diffraction, which is a result of the entrance and focal plane filter within the instrument.

To avoid saturated images as much as possible, we analyzed only every other AIA image, choosing the low-exposure image if available. To correct for the diffraction pattern, we deconvolved the AIA images with the point-spread functions provided by the instrument team. To account for flux from quiescent regions that are below the digitization threshold in the short-exposure images, we compared the flux in the quiescent region in the short-exposure images with the one in the regular-exposure images to derive an offset. The short-exposure images were subsequently corrected by this offset. The integrated flux of selected channels is presented in Fig.~\ref{Fig:aia}, with the GOES light curve for comparison. In addition, a pair of movies were created for the same time range: one of the full disk\footnote{\tiny\url{https://www.youtube.com/watch?v=w1KIncyChsE}} and another one with a smaller region-of-interest (ROI)\footnote{\tiny\url{https://www.youtube.com/watch?v=zZ9LQbL93n4}} centered on active region NOAA~12673.

\section{Results} \label{sect:results}
In this section, we discuss various methods of investigating the two flares through the HARPS-N data, without additional context as if it was another star. These begin with a look at the contrast profiles, which are then compared with the short timescale variations of activity indices, RV measurements, and EW variations.  Finally, high-resolution SST observations of the same event are presented. 

\subsection{Contrast profiles}

\begin{figure*}[t]
	
	\centering
	\includegraphics[width=\textwidth]{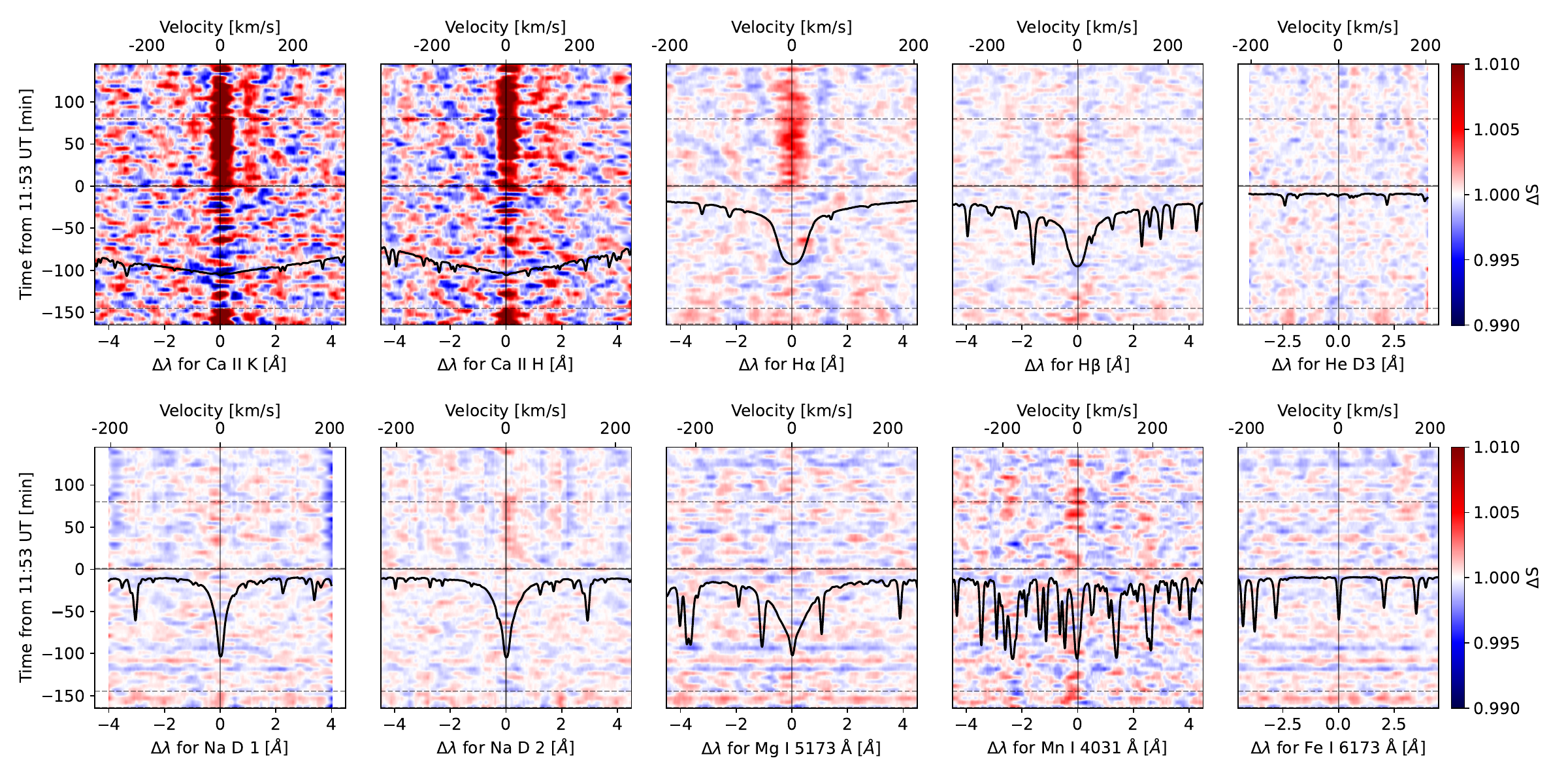}
	\caption{Contrast profiles spanning the entire HARPS-N time series of 10 spectral regions that are centered on the \CaHK, \Halpha, \Hbeta, \ion{He}{I}\,D$_3$, \ion{Na}{I}\,D$_1$ and D$_2$, \ion{Mg}{I} 5173~\AA\, \ion{Mn}{I} 4031~\AA, and \ion{Fe}{I} 6173~\AA. Each plot shows both the wavelength and Doppler velocity centered at each respective line versus the time starting from the beginning of the X9.3 flare. }
	\label{Fig:contrastprof}

\end{figure*}

\begin{figure*}[t]
	
	\centering
	\includegraphics[width=\textwidth]{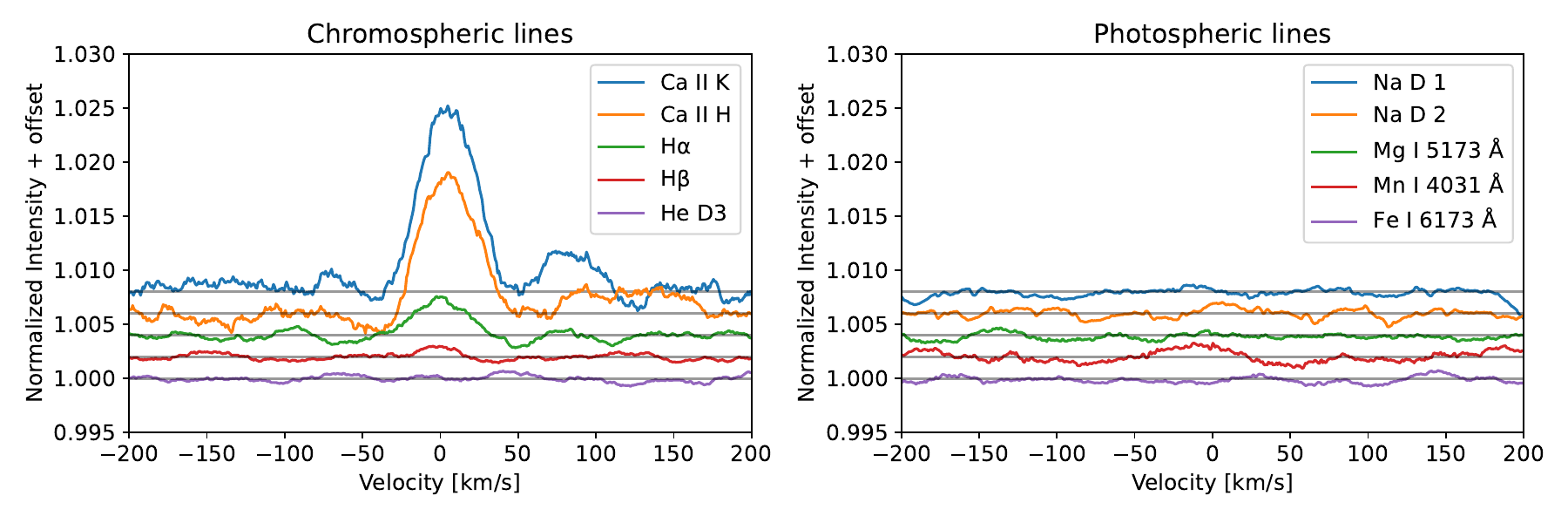}
	\caption{Integrated contrast profiles for the X9.3 flare showing each spectral window from Fig.~\ref{Fig:contrastprof} averaged over 0 to 80~min.}
	\label{Fig:activityintx9}

\end{figure*}

\begin{figure*}[t]
	
	\centering
	\includegraphics[width=\textwidth]{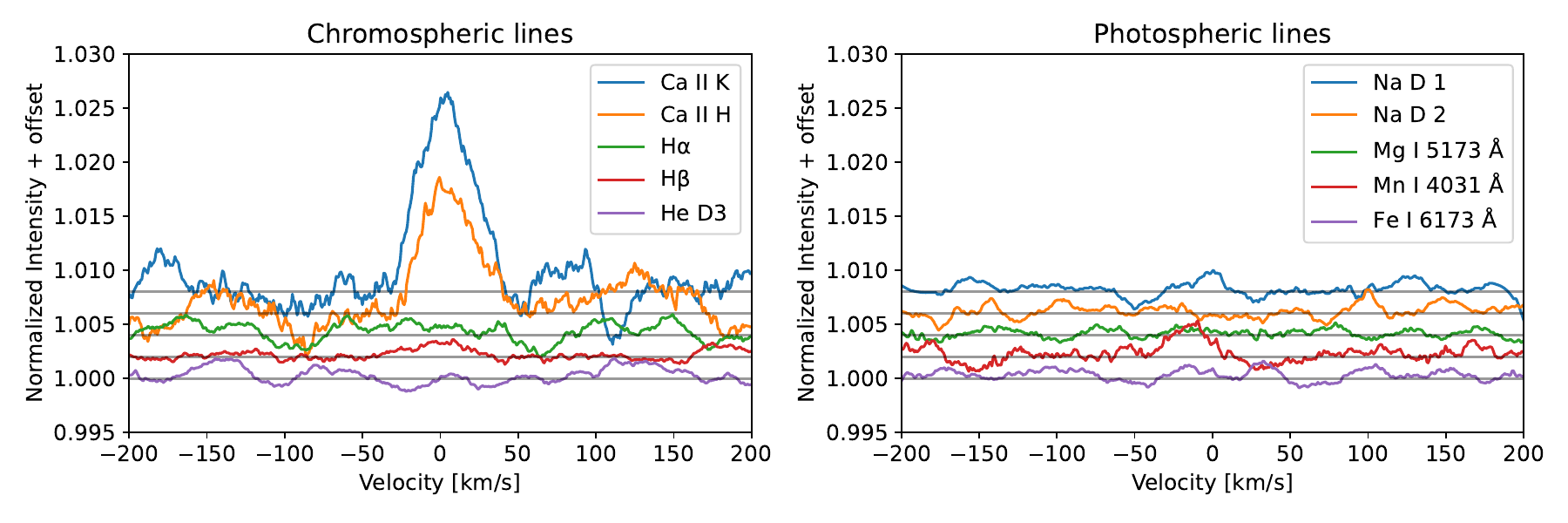}
	\caption{Same as Fig.~\ref{Fig:activityintx9} but averaged over $-164$ to $-145$~min.}
	\label{Fig:activityintx2}

\end{figure*}

In Fig.~\ref{Fig:contrastprof}, we show the so-called contrast profiles of ten selected lines from the HARPS-N spectrum. These profiles are obtained by dividing each time step by an average of the time series between 10:30 and 11:30 UT, this was done to ensure that the reference level for non-flaring spectra was taken at a time when the locations and conditions of all the solar active regions on the solar disk are as close as possible to those during the flares. This ensures that as much of the registered changes as possible are due to the flares and not unrelated evolution. Ideally one would use a longer baseline that happens before both events, but this was not available for this data set. Observations on other days were either plagued with bad weather or represented an active region geometry that was too different from the current one. 

The spectral lines in Fig. \ref{Fig:contrastprof} are shown in columns from left to right with decreasing quiet-Sun formation height \citep[e.g. ][]{Kuridze16,Sasso_2017,Jaime19,Carlsson2019}. These specific lines were picked because of their use as activity indices (see Sect.~\ref{sect:activityindex}). The two exceptions are the \ion{Mn}{i}~4031~\AA\ and \ion{Fe}{i}~6173~\AA\ lines. The latter has been added because it is the same line that is observed by HMI, which makes it one of the best-studied iron lines in solar physics, while the manganese line was selected due to its potential as an activity indicator, something which has been debated since the late 20th century. 

From this plot, we can see a constant broadening in the line core of most metrics. 
The \CaHK lines show by far the strongest response in both flares, with brightening close to 2.5\% above the non-flaring base level, as well as with a similar width. This can be seen in Fig. \ref{Fig:activityintx9} and \ref{Fig:activityintx2}, which from now on will be called the integrated contrast profiles. The integration range of these profiles is taken between $-164$ to $-145$~min and 0 to 80~min, respectively, as illustrated with a pair of dashed lines in Fig. \ref{Fig:contrastprof}. 

A weaker, and less extended response is found in most other lines, except for helium, magnesium, and iron. While no response is expected in the low photosphere where the iron line forms, it is surprising that helium, which is a widely used activity proxy (see Sect.~\ref{sect:activityindex}), does not seem to respond to the flare at all. It is possible that the flare is simply too weak, or that there is some kind of angular dependence on the response strength of this line. 

The Na line profiles both show an asymmetrical structure during both flare times, which is consistent with the findings of \citet{Rutten11}. Given the fact that this line is used regularly for exoplanet atmosphere characterization, further investigations looking at stronger flares on different stars are warranted, especially if observing on the `quiet' side of the lines could decrease the line's sensitivity to the background activity of the host star. In the next section, we explore this by redefining the sodium activity index in such a way that it deals with this asymmetry.

No signature of the flare is detected in the magnesium, even though this line has been used as an activity index \citep[e.g.,][]{Sasso_2017}. However, two strong and slightly delayed signals can be seen in manganese, which corresponds to the \ion{Mn}{I}~4031~\AA\: and \ion{Mn}{I}~4033~\AA\: lines. While no response was detected from the \ion{Mn}{I}~4034~\AA\: line. \citet{Doyle1992} reported a delayed response to a flare in these lines (their Figs.~6 and~7), in particular, a \ion{Mn}{I} line at around 4030~\AA\: that increases in intensity during the impulsive phase, and remains above the average level in the gradual phase, above the enhancement shown by most other lines around it. This is in line with our observations. However, not much else is known about this line as most works focused on the \ion{Mn}{I} 5432/5395~\AA\ lines \citep[e.g.,][]{Livingston1987, Olexa04, Vince_2005, Livingston2007, Bergemann_2019} which were shown to trace the solar cycle and seem to react to chromospheric plage. It was first suggested by \citet{Doyle2001} that their chromospheric sensitivity is due to optical pumping by \ion{Mg}{II}\,k, while \citet{sanja05,Sanja07} suggested that this line has an above-average sensitivity to solar activity. It was later shown by \citet{Vitas_2009} that this sensitivity is due to an excessive hyperfine structure, which dominates over the thermal and granular Doppler smearing that most narrow photospheric lines suffer from, effectively making it uniquely sensitive to magnetic fields. A similar test was performed on the data presented here, for these lines. No response to the flare was detected, with spectral windows looking similar to those of magnesium and iron (see Fig.~\ref{Fig:contrastprof}). 

Besides the central increase in the contrast profiles, it is also possible to make out slight increases on the red side of the \CaHK and \Halpha\ lines that overlap in \ion{Ca}{II}\,K and \Halpha but not in \ion{Ca}{II}\,H (Fig.~\ref{Fig:activityintx9}). These structures could be signs of chromospheric evaporation \citep[e.g.,][]{Kuridze_2015}, with the mismatch in the profile of \ion{Ca}{II}\,H likely being caused by the higher amount of blends on the red side \citep[e.g.][Fig. 2]{Julian15} of the line, as well as by the H$\epsilon$ line located at about 150~km~s$^{-1}$ \citep{Krikova2023}. This would imply chromospheric evaporation of around 100~km~s$^{-1}$. However, further studies would be required to corroborate this.

\subsection{Short-term activity index variations}\label{sect:activityindex}
\begin{figure}[t]
	
	\centering
	\includegraphics[width=\columnwidth]{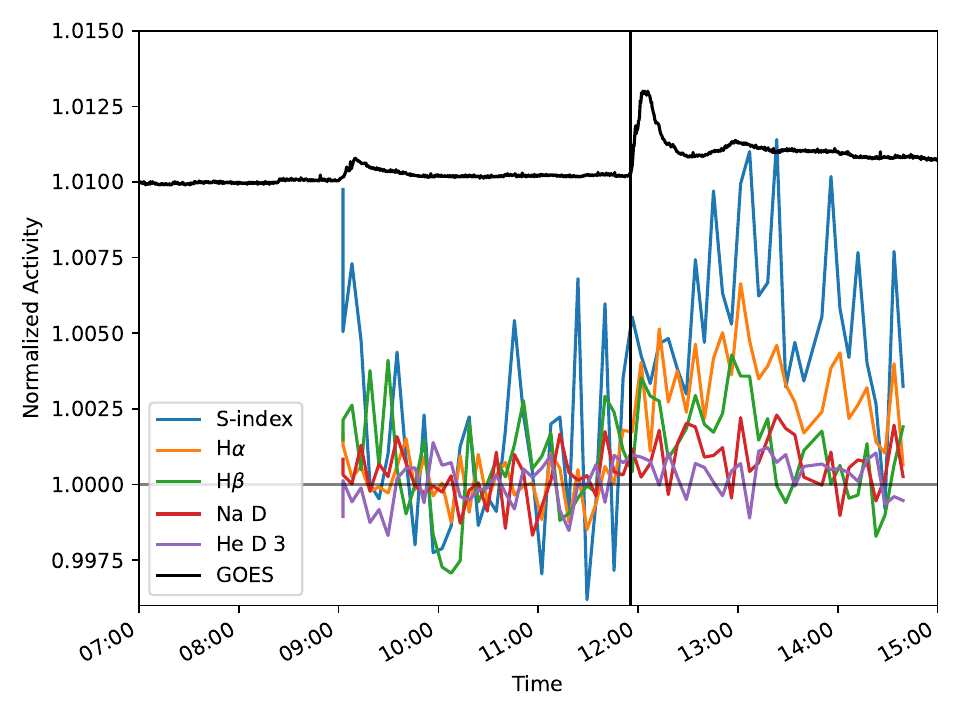}
	\caption{Relative comparison between activity indices which have been normalized to the quietest period between 10:00 and 11:00~UT. A black vertical line denotes the start of the X9.3 flare.}
	\label{Fig:activityindex}

\end{figure}

A wide variety of activity indices have been proposed over the last decades that quantify solar and stellar activity \citep[see][and references within]{Ermolli14}. By design, many of these indices are partially correlated as they react primarily to the 11-year solar cycle \citep{Schwabe44, Jrgensen2019}, or similar cycles on other stars \citep[e.g.,][]{Wilson78, Baliunas95}. However, since the lines used for deriving such indices form at different heights in the atmosphere, and therefore react differently to changes around them, one cannot expect them to be equivalent. This is illustrated by the different intensities, widths, and start times of the flare signal in the contrast profile time series shown in Fig.~\ref{Fig:contrastprof}. In addition, this gives us the opportunity to not only investigate the response of the activity indices as they are used today but to investigate if perhaps a different definition would fit the data better. 

In the case of our HARPS-N observations, the five most responsive indices have been plotted in Fig.~\ref{Fig:activityindex}. Each of these indices and the lines that they are based on are described below.  

\subsubsection{\CaHK based S-index}

The \CaHK lines centered at 3968.47~\AA\ and 3933.66~\AA\ belong to the deepest lines in the visual spectrum. Their broad wings sample the high photosphere up to the temperature minimum and can be modeled in local thermodynamic equilibrium \citep[LTE; e.g.,][]{2012sheminova}. The core, which usually has a characteristic `M-shape', forms as a result of the temperature rising in the low chromosphere, decoupling from the source function which has to be modeled in non-LTE with partial redistribution. The line core forms high up in the chromosphere below the \Mghk lines \citep{Vernazza(1981), Leenaarts2013, Bjorgen2018}. Strong variations in the line shape can be seen in spatially resolved solar observations of areas with activity \citep[e.g.,][]{Jaime13,NagaVarun2018}, while in disk-integrated observations this change expresses itself with a brightening in the cores of these lines. This change has been used as a measure of stellar and solar activity for nearly half a century in the form of the so-called S-index \citep{Wilson78, Oranje(1983b), Gomes11, 2018MNRAS.473.4326A, Dineva2022, 2023MNRAS.524.5725A, sowmya2023}, where a triangular bandpass is used to integrate the line cores, which are then weighed with a nearby pseudo continuum in the following way: 
\begin{equation}
    S = \frac{H+K}{B+V},
\end{equation}
where $H$ and $K$ refer to the integrated flux over a 1.09-\AA-wide window centered on 3968.47~ \AA\ and 3933.664~\AA, respectively. $B$ and $V$ are normally selected as 20-\AA-wide band around 3900 \AA\ and 4000 \AA, respectively. However, our HARPS-N data starts at 3900~\AA, which limits the size of $B$ to half of what is usually used. This, combined with the ghosts and low filter throughput that plagues the instrument, results in higher noise and lower absolute index values than expected. However, as these effects are largely constant over time, we can still look at the normalized activity index, where we divide the indices by the mean spectral index between 10:00 and 11:00~UT. We repeat this step for all other indices to facilitate easier comparison. 

The S-index (blue line in Fig.~\ref{Fig:activityindex}) shows strong enhancement during the X2.2 flare at the beginning of the time series. It also increases by around 1\% during the X9.3 flare, peaking at about 1~hour after the start of the flare. The peak of the S-index activity is approximately equal for both flares, although the X9.3 flare causes a much longer-lasting increase.

Between 11:30 and 11:50~UT, that is between the two flares for 20~min before the X9.3 impulsive phase, a pulsating behavior is seen in the S-index, which cannot be easily discerned in other lines, nor the integrated AIA curves (Fig.~\ref{Fig:aia}), but is present in the contrast profiles (see Fig.~\ref{Fig:contrastprof}, \CaK and \CaH panels at the times immediately before 11:53~UT). When looking at the AIA video in the 1600~\AA\ and 1700~\AA\ channels, these peaks coincide with small flare-like brightenings that do not show up in the GOES curve. Depending on the activity cycle, the cutoff for flare detection of the GOES satellite lies between low C-class to mid B-class flares \citep{Sadykov2019}, in our case the baseline being around C1 class. Thus, if the brightenings seen in the AIA data are indeed flares, they would be below C-class. However, they could also simply be due to the ubiquitous variations in activity on the solar disk. 

\subsubsection{\Halpha and \Hbeta indices}
\begin{figure}[t]
	
	\centering
	\includegraphics[width=9cm]{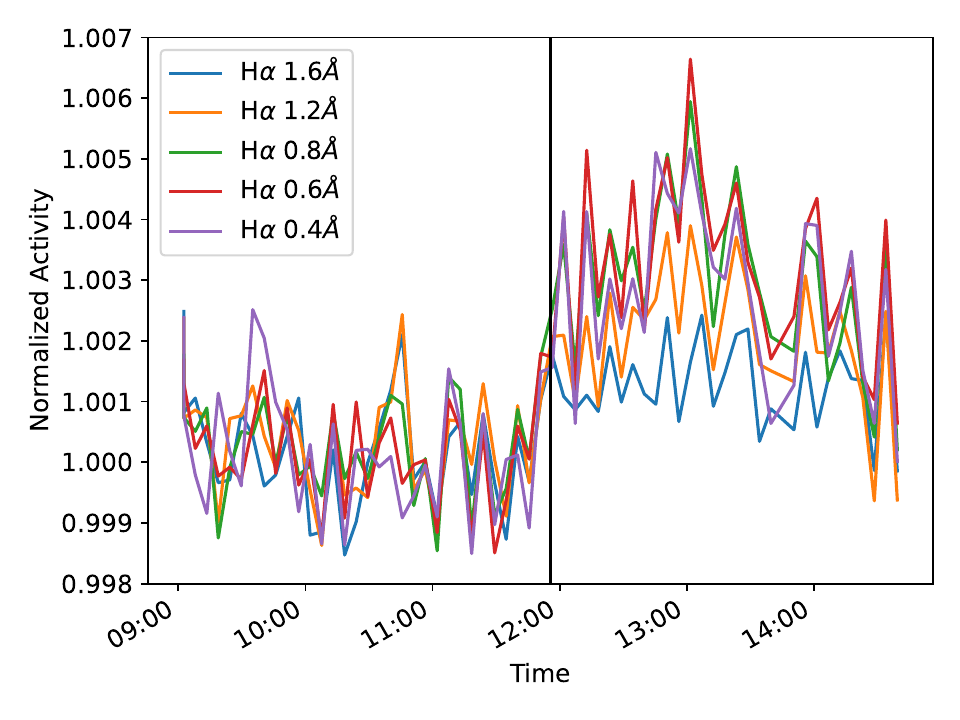}
	\caption{Variation of \Halpha core integration window for activity index.}
	\label{Fig:haactivity}

\end{figure}

The hydrogen \Halpha and H$\beta$ lines are located at $6562.79$~\AA\ and  $4861.35$~\AA\ respectively. Both lines require three-dimensional effects to be modeled \citep{2012Leenaarts}, although it has been suggested that the high mass density in the active region would make it possible to model the \Halpha line in one dimension \citep{Johan2019}.

In non-flaring conditions, both lines are in absorption and form principally as a result of the scattering of light in the chromosphere from the intense incident continuum emission produced at photospheric heights \citep{2012Leenaarts}. The \Halpha line is much more commonly studied in solar work than the \Hbeta line, which shows broadly a similar behavior with a line intensity ratio of approximately 0.3 to 0.4 when compared to the \Halpha line \citep{2017Capparelli, 2019KozaHbeta}. In quiet-Sun conditions, the line-center-to-continuum intensity ratio is sensitive to column density rather than temperature \citep{2012Leenaarts, 2015Leenaarts, 2022Druett} displaying the canopy of chromospheric fibrils in the line center and granulation in the far wings, with an opacity gap (due to the temperature minimum region of the Sun that occurs just above the photosphere) causing a rather abrupt switch between these views. The quiet-Sun \Halpha line width is correlated with typical chromospheric temperature variations \citep{molnar2019}, although this correlation is via an indirect cause, with line width being principally sensitive to column density present above the temperature minimum layer or opacity gap. Thus, under standard line synthesis, only 8\% of the line width variations are attributable to thermal broadening, that is, variations of around 0.025~\AA\ out of total observed variations of order 0.3~\AA\ \citep{molnar2019}. This opacity broadening can itself be useful to infer line-of-sight-column masses in resolved images \citep{Pietrow22}. In flares, both \Halpha and \Hbeta go into emission and exhibit extreme broadening and Doppler shifts \citep{Ichomoto1984, wuelserr1989, DruettPoster, Pietrow22}. This occurs over large sections of the flare ribbon but is particularly acute in small `flare kernels' associated with energetic events such as high-energy beam particle heating \citep{2017Druett, Zharkov_2020, 2022Osborne}. 

Although such metrics as strong red-wing enhancements are highly valuable in spatially resolved observations of solar flares, the most effective bandwidths for detecting flare activity in stellar observations is still a matter of debate, usually framed in particular implementations of the \Halpha activity index \citep[and references within]{GomesdaSilva2011}. Similarly to the S-index, this index takes the average over the \Halpha line core and divides it by two sections of pseudo-continuum to either side of it. \citet{Bonfils07} and \citet{Boisse09} define it as
\begin{equation}
    \mathrm{Index} = \frac{F_{\mathrm{\tiny\Halpha}}}{B+V},
\end{equation}
where $F_\mathrm{\tiny\Halpha}$ corresponds to a 0.6-\AA-wide window centered at 6562.808~\AA, and $B$ and $V$ are both 8-\AA-wide windows centered around 6550.85 and 6580.28~\AA, respectively. 

Not long after this definition, a larger width of 1.5~\AA\ has been used for $F_{\tiny\Halpha}$ \citep[e.g.,][]{Cincunegui07, meunier2009, Gomes11}, as very strong activity can broaden the line core to this point and beyond \citep{Pietrow22} and the signal-to-noise ratio of a larger band is of course better. However, when averaged over the full disk, this bandpass includes a significant photospheric contribution \citep[e.g.,][Fig.~1]{Watanabe2011} which washes out the chromospheric signal in all but the most extreme cases. A similar conclusion has been drawn by \citet{gomes22}, who reported that the narrower bandpass correlates more strongly with the S-index. We confirm this, as using the broader bandpass washed out almost entirely the signal of both flares, while the narrower bandpass tracks better with the S-index. In Fig.~\ref{Fig:haactivity}, we show a series of normalized activity indices with differing bandpass values, showing the wash-out effect of large bandwidths.  

We take a similar approach for \Hbeta, where we define a bandwidth of 0.2~\AA, which was chosen to cover a similar fraction of the line core as with \Halpha (0.4~\AA), and later narrowed down to compensate for the lower opacity of the line. For the $B$ and $V$ bands, we use the definitions by \citet{West2008} which span a 10-\AA-wide region centered around 4845 and 4880~\AA, respectively. 

The orange line, which represents the \Halpha activity index in Fig.~\ref{Fig:activityindex} is much less noisy than the S-index. This is likely due to the higher throughput of this section of the HARPS-N bandwidth, as well as the lack of any ghosts, which are present in the \CaHK window. The flare causes a roughly 0.5\% increase in activity around the time of the X9.3 flare and increases several minutes before the onset of the flare as discussed in \citep[e.g.,][]{Benz2016}, and tracks the GOES curve more closely than the S-index. The smaller X2.2 flare does not leave a significant imprint on the index.

The green line, which represents the \Hbeta index, peaks near the X2.2 flare, showing more sensitivity for it than \Halpha. For the X9.3 flare, we see an increase of similar magnitude as for the smaller flare, but with a longer duration. However, the increase declines much faster than for \Halpha.

Similarly in behavior to the \CaHK lines, \Halpha and \Hbeta brighten just before the GOES SXR curve indicating the start of the flare at 11:53~UT (see Fig.~\ref{Fig:contrastprof}), with offsets on the order of 5~min between these signals reported in other flares \citep{2023Singh_Riblets} and also in historical literature \citep{1974Kane}. The response is stronger in \Halpha than in \Hbeta, which \citet{Zirin1982} noted and attributed to the higher opacity of the line and \citet{2017Capparelli} used to infer electron beam statistics using RADYN models \citep{2015Allred}. The \Hbeta line has a shorter response time with emission decreasing from the peak values back down to quiescent levels about an hour earlier than \Halpha, which could be attributed to the lower enhancement of \Hbeta above the background emission strength compared to \Halpha, as well as potential differences in the formation processes of the lines. 

\subsubsection{Na index}
\begin{figure}[t]
	
	\centering
	\includegraphics[width=9cm]{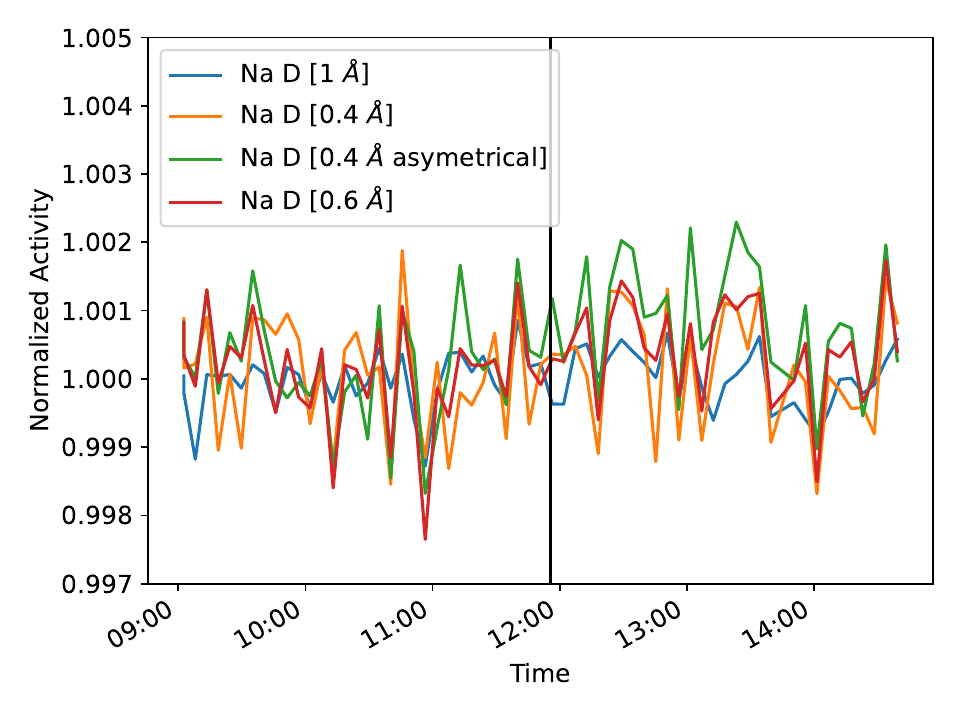}
	\caption{Variation of \ion{Na}{I}\,D$_1$ line integration window for activity index.}
	\label{Fig:naactivity}

\end{figure}
The \ion{Na}{I}\,D$_1$ and $D_2$ lines centered at 5895.92~\AA\ and 5889.95~\AA\ are strong resonance lines that require three-dimensional NLTE modeling \citep{Leenaarts10,2011lind}. 
These lines are often described as purely chromospheric \citep[e.g.,][]{Gomes11} but over the last few years it has been shown that the line core of these lines is sensitive to a range that goes from the high photosphere to the low chromosphere \citep{Kuridze16}. In general, for the quiet Sun, the lines form just above the temperature minimum, while the formation height moves down significantly in areas with high magnetic concentrations \citep{Leenaarts10}, meaning that the line core brightness correlates to magnetic bright points in the lower photosphere. It has also been shown by \citet{Rutten11} that the \ion{Na}{I}\,D$_1$ line suffers from an asymmetry in formation height, with the blue wings showing inverse granulation, while the red wings show normal granulation lower down. In the contrast profiles, we see that the opposite is true for the \ion{Na}{I}\,D$_2$ line, where the asymmetry is on the blue side. Despite this, \citet{Short1998} showed that these lines do track the \Halpha index well. 

The two lines are used together for a single activity index in a similar fashion as for the S-index.  \citet{Diaz07} defines
\begin{equation}
    \mathrm{Index} = \frac{D_1+D_2}{B+V},
\end{equation}
where $D_1$ and $D_2$ correspond to two 1-\AA-wide windows centered on the two sodium lines. $B$ and $V$ are in this case 10- and 20-\AA-wide windows centered at 5805.0 and 6090.0~\AA, respectively. However, instead of taking the average of these windows, the 10 highest flux values inside of these bands are averaged. 

Similarly to the \Halpha index, this band is very wide and samples photospheric contributions \citep[e.g.,][Fig.~1]{Rutten11}. In addition, the response of the line is asymmetric, so we adapt our window to match both of these aspects and end up with a 0.4-\AA-wide window centered 0.1~\AA\ bluewards and redwards of the respective line cores. In Fig.~\ref{Fig:naactivity}, we show the difference between the index as defined by \citet{Diaz07}, a symmetrical index with a narrower width, and the one described above. Both flares do not show any response to the classical index, while the strongest signature can be seen in the asymmetric index. 

While it is hard to call this a detection, the red line in Fig.~\ref{Fig:activityindex} is slightly raised with respect to the mean value from the moment when the X9.3 flare starts, especially when compared to the broader indices shown in Fig.~\ref{Fig:naactivity}.

\subsubsection{He index}
The \ion{He}{I}\,D$_3$ line at 5875.62~\AA\ requires to be modeled in three-dimensional NLTE \citep{Libbrecht2021}. Due to its low opacity, this line is dominated by the photospheric continuum in quiet regions, while it gains opacity in active regions \citep{Landman81, Heinzel2020}. This means that in quiet-Sun environments, the line is hardly present, that is, it appears and becomes deeper as more activity is present \citep[Fig.~1]{Landman81}. For this reason, it is necessary to correct for the continuum component when studying this line in resolved observations \citep[Figs.~4.8 and~4.9]{Tine2016}. In addition, we can see from the same figure, as well as those presented in \citet{Libbrecht2021} that this method could potentially be utilized with activity indices, where the index is divided by a quiet day or the outer wings of the profile. However, the S/N ratio of our data is not sufficient to meaningfully test this idea. Therefore, we used the classical He activity index as defined by \citet{Boisse09}.
\begin{equation}
    \mathrm{Index }= \frac{F_{\tiny\mathrm{He}}}{B+V},
\end{equation}
where $F_{\tiny\mathrm{He}}$ corresponds to two 0.4-\AA-wide windows centered at the \ion{He}{I} line, and $B$ and $V$ correspond to two 5-\AA-wide windows centered around 5869.0 and 5881.0~\AA, respectively.

No signature of either flare shows up in this index, which could be a matter of the flares being too weak, or that the flares are located so close to the limb, where higher atmospheric layers get more opacity.  

\subsection{Radial velocity measurements}
The RV method detects planets by observing the host star's reflex motion-induced velocity changes, which are typically of the order of a few meters per second for mini-Neptune exoplanets, but a few tens of centimeters per second for Earth-like ones. This technique measures the spectral shift compared to a reference, or as a function of time to find periodicity. However, besides traditional noise sources due to the instruments \citep{Dumusque(2021), Cretignier(2021)}, such detections are made more difficult by variations in activity on the stars themselves \citep{Meunier-2010a, Dumusque-2011a, Meunier:2017ab, Cretignier(2022), Sen23}. Indeed, the transit of active regions across the stellar disk has been known to cause signals in RV similar to those of planets \citep[e.g.,][]{Queloz01, Bonfils07, Huelamo08, Meunier-2010a, Dumusque-2014b, Zhao:2023aa}. As a consequence, several planetary claims have been shown to be induced by stellar activity or are still debated today, for instance, the case for Barnard's star \citep{Ribas(2018), Lubin(2022)}, Kapteyn's star \citep{Anglada(2014), Robertson(2015), Ji(2019)}, and HD\,41248 \citep{Jenkins(2013), Santos(2014), Jenkins(2014), Feng(2017b), Faria(2020)}.

Since the RV value is obtained from broad bandpass high-resolution spectra, the RV observations are often paired with monitoring of one or more activity indices, which can be used to retrieve the rotational period of the star \citep{Butler2017, Langellier2021}, especially on stars where active regions can persist for hundreds of rotations \citep{Robertson_2020}. However, the sparse observational sampling of RV stellar observations can provide a poor temporal sampling for active regions that evolve rapidly while on the disk, or transient events such as flares that can still cause RV offsets of a few meters per second on M-dwarfs in the infrared \citep{Reiners_2009}, this is also called `activity jitter'. Additionally, simulations have shown that the contribution function of photospheric lines can expand well into the chromosphere where peak beam energy deposition occurs \citep{Monson2021}, confirming that a strong enough flare should have an RV imprint. 

\begin{figure}[t]
	
	\centering
	\includegraphics[width=9cm]{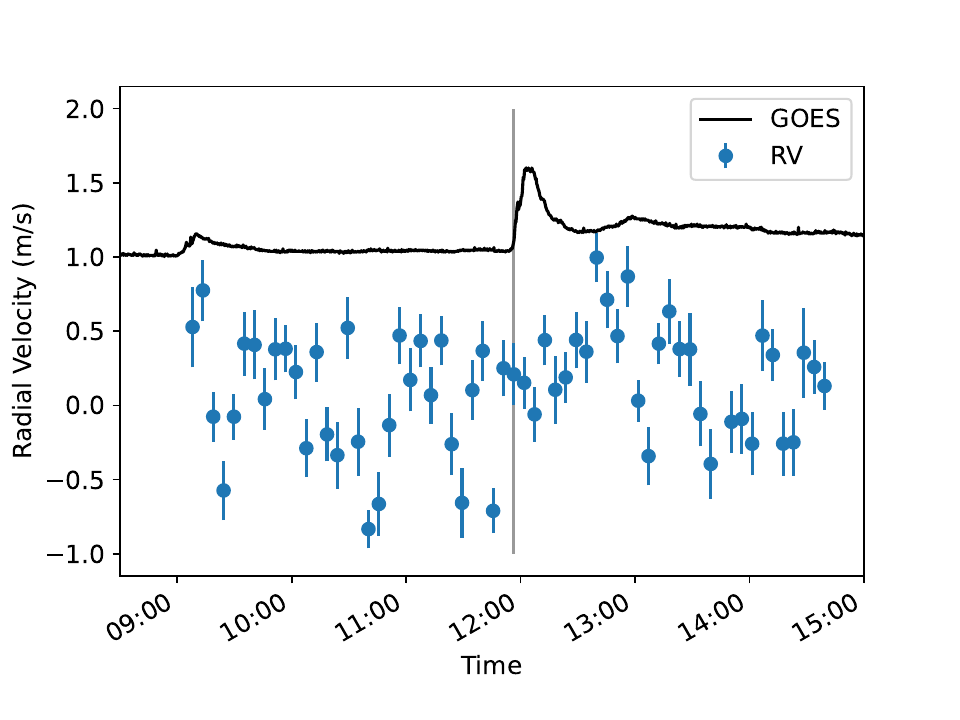}
	\caption{Relative comparison between activity indices which have been normalized to the quietest period between 10:00 and 11:00~UT. The gray vertical line denotes the start of the X9.3 flare.}
	\label{Fig:rv}

\end{figure}

We investigate the effects of these flares on solar RVs. The RVs were extracted using a cross-correlation function \citep[CCF,][]{Baranne-1996, Pepe-2002a} with a tailored line selection of 3277 photospheric stellar lines between 3900~\AA\ and 6828~\AA. Lines are selected based on morphological criteria of the line profiles in order to avoid the strongest blends \citep{Cretignier(2020a)}. Only lines presenting a clear line profile with a resolved line core were kept. The use of a CCF in signal processing is a typical way to extract the small shift signatures by agglomerating the information of thousands of photospheric stellar lines. The RV curve was then de-trended by the daily averaged RV to remove the effect of active regions on the solar rotation. The de-trended RVs with their corresponding error bars were plotted in Fig.~\ref{Fig:rv}, together with a scaled GOES curve. 

Due to the instrumental stability and granulation signals, we do not expect an RV dispersion smaller than 50~cm~s$^{-1}$ \citep{AlMoulla(2023)}. While the scatter in the points is too large to claim a detection, a tightening of the scatter and a small bump of a few tens of centimeters per second do coincide with the GOES curve, although the second peak corresponds to the secondary bump rather than the X9.3 flare peak in much the same way as the S-index did. 
Even if the present work cannot significantly detect the signature in RVs, an upper limit of 50~cm~s$^{-1}$ can be fixed which represents the intrinsic stability of the HARPS-N spectrograph. This value is lower than the one obtained in \citet{Reiners_2009} which can be explained either by the different flare intensities, the different stellar spectral types, or the different spectral ranges of the spectrographs.

We then studied in a similar fashion the RV time series of each individual line rather than the average value obtained from the CCF. The line-by-line (LBL) RVs were obtained as in \citet{Dumusque(2018)} using a template matching method.  The vast majority of the LBL RVs were not correlated to the GOES peak. This is not surprising given that the LBL RVs precision does not exceed $\sim$5~m~s$^{-1}$ for the best lines. A closer investigation of the top three most correlated lines reveals that they are all blends (4947.6~\AA: \ion{V}{II} and \ion{Si}{I}, 5664.0~\AA: \ion{Ni}{I} and \ion{Cr}{I}, and 4327.1~\AA: \ion{Fe}{I} and CH). Previous studies have shown that RVs coming from blended lines should be carefully interpreted \citep[Appendix]{Cretignier(2020a)}. This raises the question if a similar, and more significant, trend would emerge when investigating a much stronger stellar flare with the same methods.  

\subsection{Equivalent width measurements}


Recent high-resolution spectroscopic studies have demonstrated how the EW of some atomic lines increases with stellar activity. Different studies have confirmed that spectral lines are systematically deeper in active stars than in quiet ones with similar stellar parameters. \cite{2017reddy} demonstrate that iron lines forming in the upper layers of the photosphere (that is, at small optical depth) yield larger abundances than those forming deep in the photosphere. The authors also show how the barium over-abundance correlates with stellar activity in young stars. Following this, \cite{2019galarza} shows a similar behavior of spectral lines when analyzing the young ($\sim$400~Myr) solar-analog HIP36515.
The stellar parameters derived with the classic spectroscopic equilibrium approach vary with activity as well, as a consequence of this activity-related enhancement of the spectral lines. Significant changes of effective temperature $T_{\mathrm{eff}}$, surface gravity $\log g$, and (in particular) micro-turbulence velocity parameter $\xi$ are detected along the activity cycle, which translates also in changes of the overall iron content [Fe/H] and age determination. 
Later, \cite{2020spina} expanded the analysis to a sample of 211 Sun-like stars and investigated the behavior of 20 different atomic elements. The authors report an increase of EW with activity along the stellar cycle especially in moderate to strong lines. This is in line with  \cite{2019galarza} and \cite{2020spina}, who report that the major effects are visible in lines with EW $\gtrsim$ 50\,m\AA\ or optical depth $\log \tau < -1$ dex. As a consequence of the spectroscopic abundance analysis, these effects result in an over-estimation of $\xi$ and under-estimation of $T_{\mathrm{eff}}$, $\log g$ (even though to a smaller extent), and [Fe/H]. For Ba, the authors confirm the previous study by \cite{2017reddy}.  Finally, \cite{2020baratella_ges, 2021baratella} analyse dwarf stars in several young open clusters (OCs) and reach the same conclusions. On the one hand, they state that the classic EW method (that is, using only iron lines, which form on a wide range of optical depth) fails when applied to very young and/or active stars. One way out is the application of new methods and a more refined selection of the spectral lines used in the analysis. On the other hand, the barium over-abundance is strictly related to the activity but the same could be said also for other elements, such as yttrium if the abundance is derived using very strong spectral lines. This is in stark contrast to solar abundance studies where typically only quiet-Sun observations are considered and where the phase of the solar cycle is known \citep[e.g.,][]{Bergeman21,pietrow23b}.



In high-resolution solar observations, it is possible to confirm interpretations of the changes of spectral lines as a result of activity \citep{Ichomoto1984, Hong2018}, for example, by using spectroscopic inversion codes \citep[e.g.,][]{Kuridze2018, Jaime19, pietrow2020, sepideh20}. However, these effects have only been studied for a select number of lines, and not in a disk-integrated environment. Therefore, it is still largely unknown what the main cause(s) of these EW enhancements is (are), although the most affected lines seem to be those that form in the highest part of the photosphere. 

The set of HARPS-N observations of solar flares offers a unique opportunity to check if we can detect the same effects observed in younger and more active stars. We measured the EW of the spectral lines listed in \cite{2020baratella_gaps} with the code ARES~v2 \citep{2015sousa}, which allows us to measure them in a fast and automatic way. The line list is specifically designed to study Sun-like stars observed with HARPS-N and it comprises a total of 225 lines of 11 different atomic species. In particular, iron lines are used to derive the stellar parameters by imposing the excitation (for $T_{\mathrm{eff}}$) and ionisation (for $\log g$) equilibria, and by imposing that the abundances of each line do not correlate with the reduced EW (for micro-turbulence velocity $\xi$). To do this, we used the automatic software \textit{qoyllur-quipu} \citep[q$^2$,][]{Ramirez2014}. 

In general, we detect no significant variations of the $T_{\mathrm{eff}}$, $\log g$, and $\xi$ with the flare detection. The scatter of the measurements along the whole time series is of the order of 70~K, 0.1~dex, and 0.15~km~s$^{-1}$ for the three parameters, respectively. These values are comparable with the typical uncertainties found with this method. We can conclude that there is no distinct signal of the flare in the stellar parameters determination via the EW method. 

At a deeper investigation, the EW of each spectral line of the different atomic species shows no variation along the time series as well. We do not see any variation of EW directly linked to the flare in any of the 225 spectral lines in the list. The scatter of the measurements of each individual line is small (within $\sim$5~m\AA\ for the majority of the lines), and mainly due to continuum placement. In \citet{2020spina} and \citet{2020baratella_ges}, variations larger than 10~m\AA\ are expected, especially in moderate and strong lines: we do not observe this in our set of spectra. This means that activity, which is not visible in the RV or white-light light curves, should not affect measurements of this kind.  



\subsection{High-resolution flare observations}

\begin{figure*}
\centering
    \includegraphics[width=0.8\textwidth]{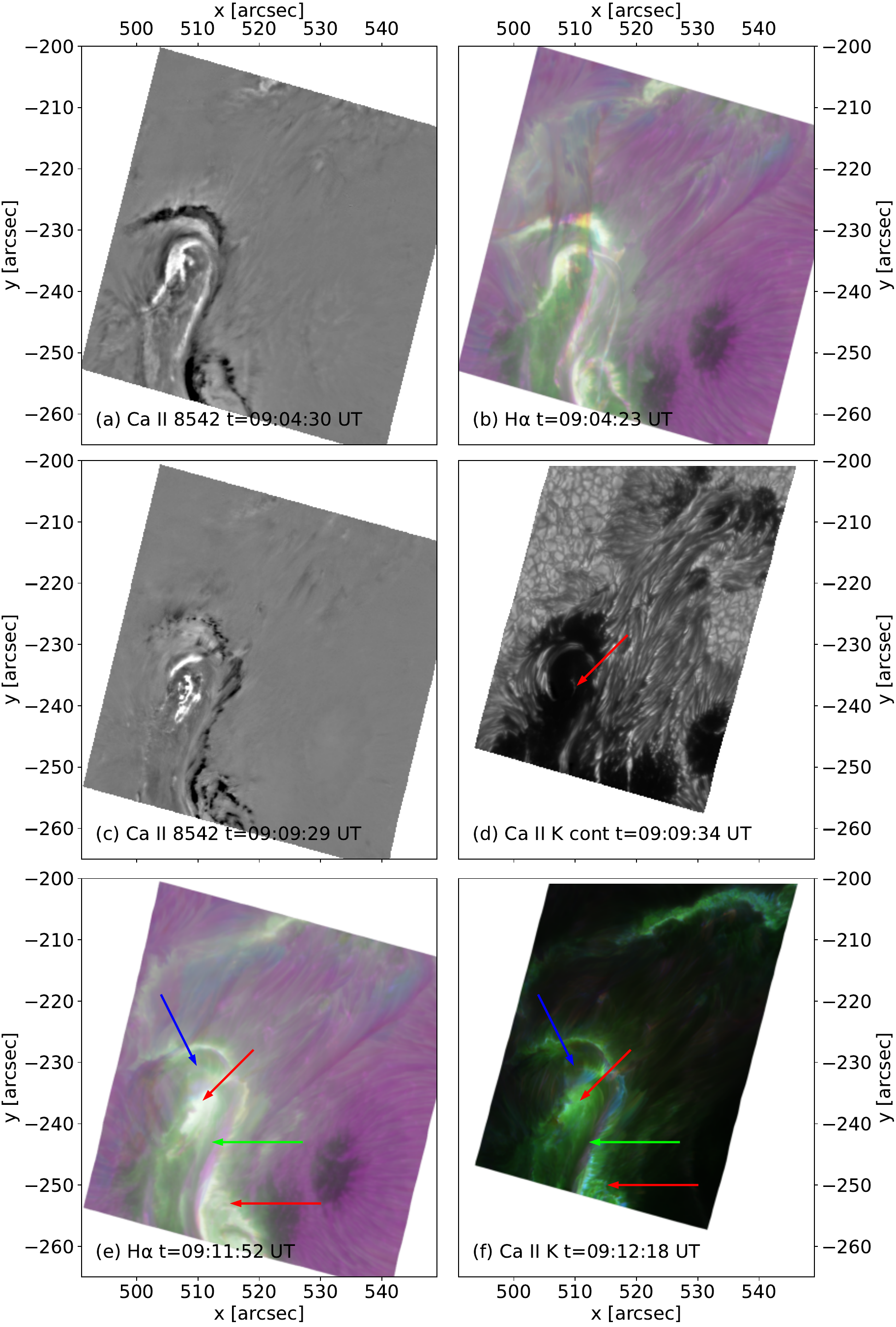}    
    \caption{Temporal evolution of the X2.2 flare in active region NOAA~12673 on 6~September 2017. (a) The early stage of the flare in the total unsigned circular polarization (Stokes-V) of the \CaIR line, showing the chromospheric imprint of the parasitic negative-polarity patch (black) impinging onto the positive patch (white) at heliocentric coordinates $(510\arcsec,\, -240\arcsec)$. (b) The ribbons at the same time in a logarithmic \Halpha COCOPLOT. (c) The expanding flare ribbons and rotation of the parasitic polarity umbra. (d) The white light response in the \CaK continuum is indicated by a red arrow. (e) A logarithmic \Halpha COCOPLOT near the peak of the flare, showing the expanded flare ribbons (red arrows), the formation of a small arcade to the left of the rotating umbra (blue arrow), and a filament (magenta) that runs from north to south between the flare ribbons (green arrow) and (f) the corresponding \CaK COCOPLOT.}
    \label{fig:X22}
\end{figure*}

\begin{figure*}
\centering
    \includegraphics[width=0.75\textwidth]{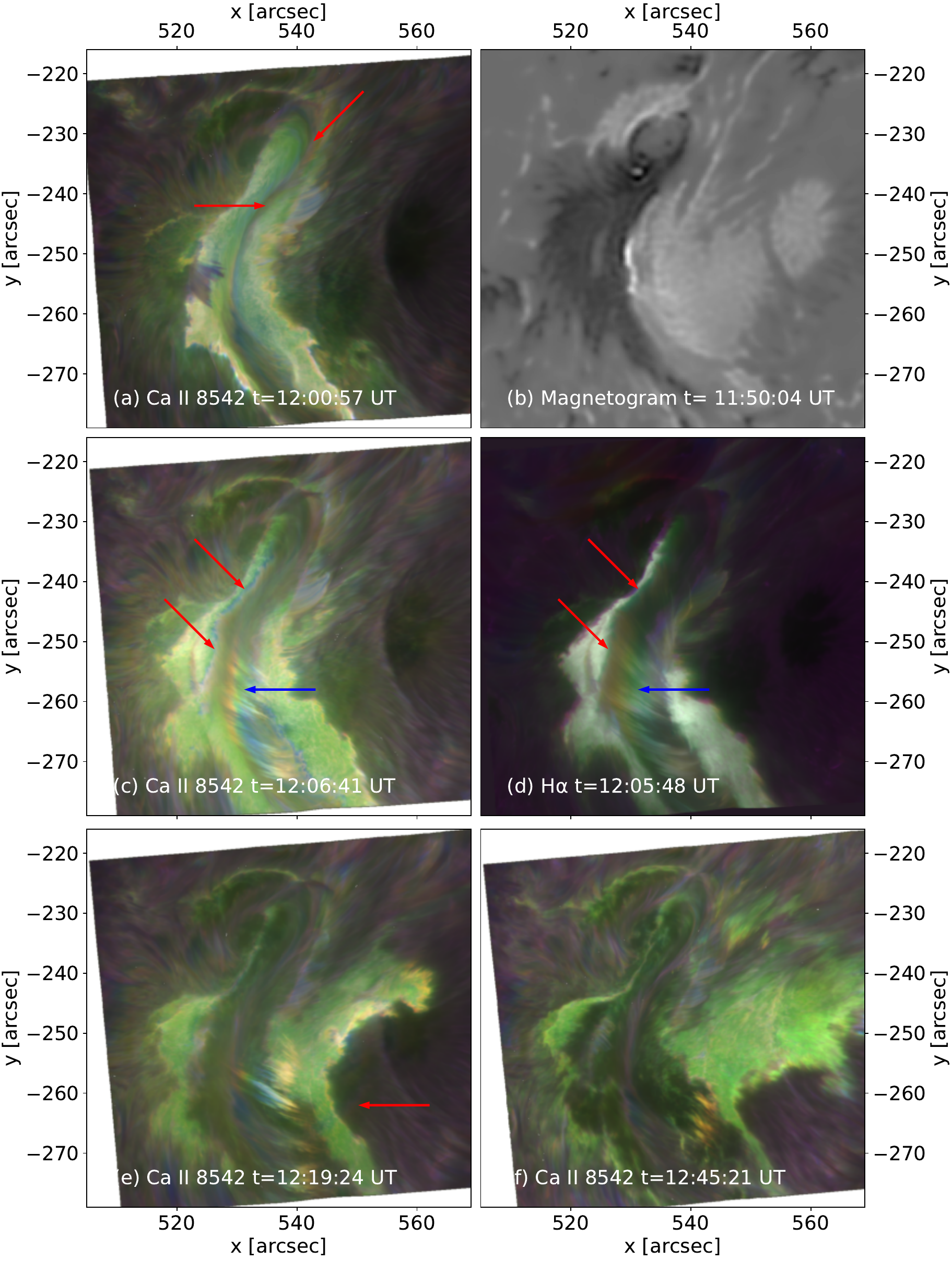}
    \caption{The magnetic context and flare ribbon evolution of the X9.3 flare in active region NOAA~12673 on 6~September 2017. (a) Two long flare ribbons near the start of the flare are shown in a \CaIR COCOPLOT. (b) The magnetic context of the flare shows an intrusion of a parasitic, negative-polarity region (\textit{left}) into the northern portion of the positive-polarity region (\textit{right}). This image was processed by the convolutional neural network presented in \citet{Carlos19}. Co-temporal (c) \CaIR and (d) \Halpha COCOPLOTs showing the initial expansion of the flare ribbons and the formation of the loop arcade. (e) A sunquake visible in the \CaIR COCOPLOT occurs before (f) the dramatic further expansion of the flare ribbon to the right of the image.}
\label{fig:X93}
\end{figure*}

 \begin{figure*}[h!]
	
	\centering
	\includegraphics[width=0.7\textwidth]{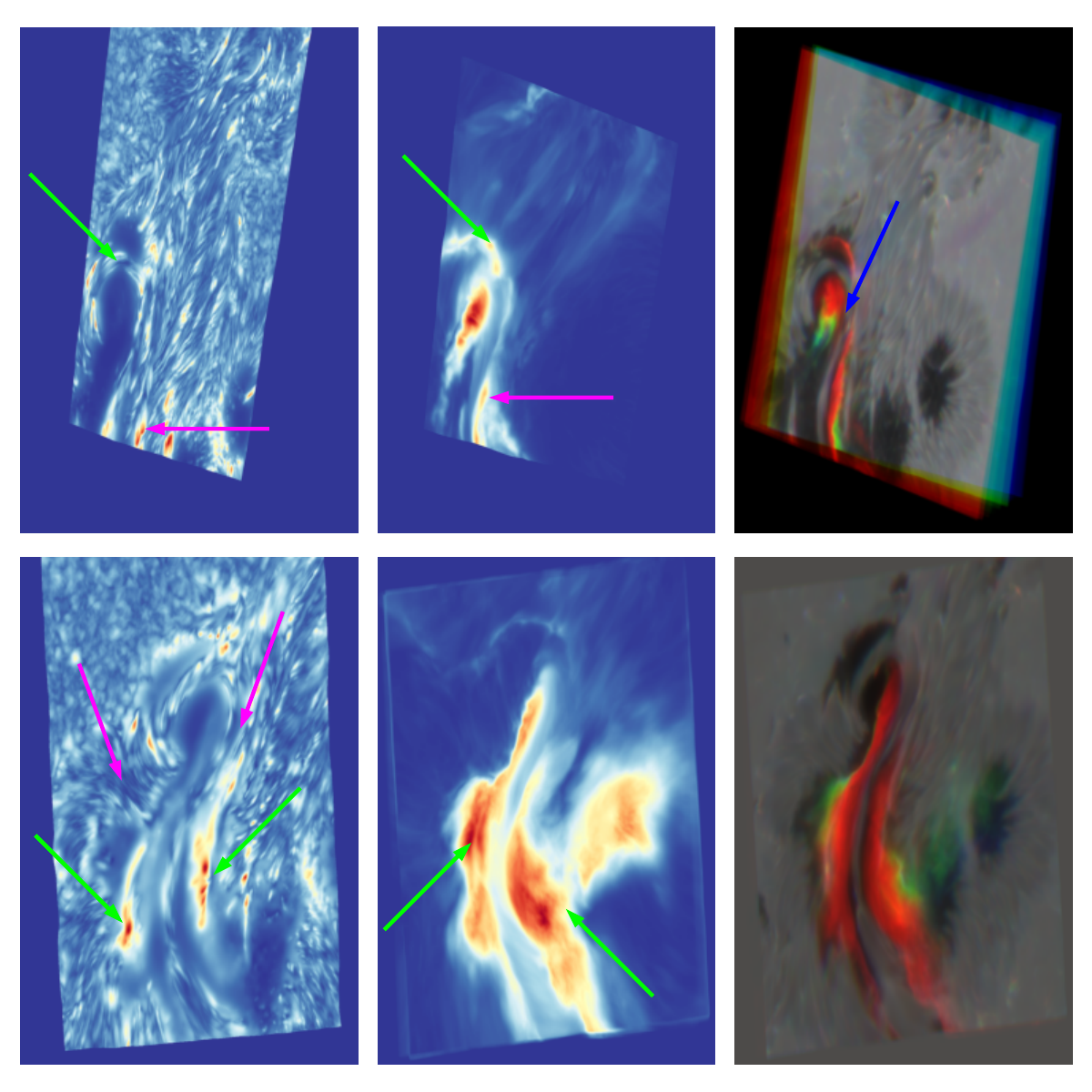}
	\caption{BaSAMs and temporal COCOPLOTs of the X2.2 and X9.3 flares mapping the points of largest change in the FOV. The first column shows a BaSAM of the 4000~\AA\ continuum point over the entire time range. The second column does the same but for the \Halpha line core. The third image captures the time evolution in the \Halpha line core, with red being early in the time series, green towards the middle, and blue late.}
	\label{Fig:bsams}

\end{figure*}

In Figs.~\ref{fig:X22} and \ref{fig:X93}, we present the evolution of the two flares in the \Halpha and \CaIR lines using Color Collapsed plots \citep[COCOPLOTs,][]{2022druettcoco}, as well as by showing the total circular polarization (Stokes-V), and a continuum image. COCOPLOTS is a technique where a three-dimensional data cube is collapsed into a two-dimensional image, processing the collapsed dimension into an RGB color value via three filters that can be used to select ranges of interest in the collapsed dimension. Here, we collapse the spectral dimension by applying the red filter to the red wing of the line, the green filter to the line center, and the blue filter to the blue wing. In the third column of Fig.~\ref{Fig:bsams} a COCOPLOT applied to the time series of the line core is shown for both flares, in order to capture the evolution of the flare in a single RGB image. 

This is combined with two background-subtracted solar activity maps \citep[BaSAMs, ][]{Denker2019b}. The concept of BaSAMs was initially introduced by \citet{Verma2012a} to monitor changes in the magnetic field around decaying pores, and more recently, \citet{Kamlah2023} presented BaSAMs for high-resolution images. This technique involves calculating an average two-dimensional map for the entire time series, subtracting this average map from each individual image, and computing the modulus of these difference maps to obtain the final average two-dimensional BaSAMs.

In the top row of Fig.~\ref{Fig:bsams}, we see a BaSAM for the X2.2 flare in the 4000~\AA\ continuum, the \Halpha line core, and a temporal COCOPLOT of the \Halpha line core. The bottom row shows the same thing but for the X9.3 flare. 

For the first flare, the most significant changes in both the continuum and \Halpha BaSAMs are observed in the curved penumbral filaments encircling the rotating umbra (green arrows), which shows the increasing helicity of the flare. In addition, noticeable small-scale changes are seen along the elongated penumbra and the lower part of the PIL, which seem to correspond to the footpoints of the flare (magenta arrows). The temporal COCOPLOT shows that the structure over the rotating umbra becomes narrower over time, and expands further outwards (blue arrow). 

For the eruptive X9.3 flare, extended regions exhibit significant signals in the BaSAMs. The footpoints of the flare loop arcades are visible in the continuum BaSAMs (green arrows), along with small-scale changes throughout the rotating umbra and along the elongated PIL (magenta arrows). In \Halpha BaSAMs, the two extended flare ribbons are clearly delineated (green arrows). Moreover, strong variations are observed in the neighboring umbra (magenta arrow), indicating the broader extent of the X9.3 flare. The COCOPLOT captures the expansion of the flare ribbons over time, as well as the acceleration of this process. 
The difference in the extent of variations in both metrics also suggests that the first flare is confined, while the subsequent flare is eruptive in nature.

In addition, the secondary peak in the GOES spectrum occurs during a period of dramatic expansion of one bright flare ribbon over the dark umbra beneath (Fig.\ref{fig:X93}e,f), and also the formation of a new arcade system outside of the SST field of view (see AIA videos).

\section{Discussion}\label{sect:flare}
The two flares, that have been investigated in this work, are amongst the best-studied solar flares ever, making them excellent candidates for a comparative Sun-as-a-star study where high-resolution observations are compared to one-dimensional disk-integrated spectra. In order to facilitate this, we break down this section into a stellar part, where only disk-integrated information is considered, a solar part where the high-resolution aspects are explored, and a final solar-stellar part where both aspects are combined. 

\subsection{Stellar context}
Compared to stellar observations \citep[e.g.,][]{Maehara2015, Pietras2022}, the observed solar flares are very weak, despite the second flare being amongst the strongest recorded on the Sun. Although the GOES curve in Fig.~\ref{Fig:VIRGO-GOES} exhibits clear peaks for both flares, no significant increase is evident in the TSI time series. This explains why a confident TSI detection of a flare is barely possible for even exceptionally large flares (larger than X10) as shown in \citet{2006JGRA..11110S14W}. The first solar flare measurement in TSI was made in 2003 after 25 years of space-born irradiance observations \citep{2004AAS...204.0215K}. The event was an X17 flare, measured by the Total Irradiance Monitor \citep[TIM,][]{Kopp2005} on NASA's SOlar Radiation and Climate Experiment \citep[SORCE,][]{Rottman05} satellite. \citet{2010NatPh...6..690K} demonstrate how difficult it is to detect a solar flare signal in TSI, even for an X-class flare. In their work, the signal of 42 flares in the range from X1.3 to X10 was combined to recover a detectable signal. For weaker flares, this is even more complex because 1477 flares in the range from M1.6 to C4 had to be averaged to achieve a significant detection. 

This places the two studied flares, with estimated energies of $2.2 \times 10^{31}$~erg and $9.3 \times 10^{31}$~erg, amongst the smallest stellar flares ever detected. This is because many stellar flare observations are made with comparatively broad filters which require a strong white-light signal to create a detectable peak. However, a significant signal is found primarily in chromospheric spectral lines, and the activity indices derived from them. 

While only a select number of works have reported flare observations with high-resolution spectra, a pattern in the contrast profile seems to emerge when these are compared to the results of \citet{Otsu_2022}. For example, a superflare on YZ~Canis~Minoris reported by \citet{Namizaki2023} seems to be located close to the limb when its contrast profile pattern is compared to Fig.~4 of \citet{Otsu_2022}. On the other hand, the superflare on HK~Aqr reported by \citet{Gonzalez2022} has been assigned a heliocentric latitude of around 34 degrees by the authors, which places it relatively close to the disk center. The contrast profiles shown in their Fig.~4 are similar to the pattern shown in Fig.~2 of \citet{Otsu_2022}, which is also taken for a flare close to the disk center, further validating the idea that the shape of the contrast profiles depends on the location of the flare on the disk. From these works, a constant shape in the profile would be expected for our flares, as they took place close to the limb. This is indeed the case. 

\textit{Activity indices and contrast profiles.}
The strongest signal was found in the \CaHK lines, where the enhancement in both the S-index and the contrast profiles was above 1\%. However, the intensity and width of these metrics were comparable for both flares, including the magnitude of the chromospheric evaporation, with the only differentiating factor being the duration of each flare. 

Our analysis extended to the behavior of the \Halpha and \Hbeta lines, revealing a weaker but still distinguishable signal if a modified narrower definition of the activity index was used. A stronger response was seen in \Halpha than in \Hbeta, likely due to the higher opacity of the former. In addition, these lines brightened several minutes before the start of the flare, which is likely a manifestation of the \citet{Neupert1968} effect (see next section). This makes these lines potentially valuable as short-term flare predictors. 

We studied multiple other activity indices, including \ion{He}{I}~D$_3$, the \ion{Na}{I}\,D lines, \ion{Mg}{I} 5173~\AA, \ion{Fe}{I} 6173~\AA, and \ion{Mn}{I} 4031~\AA. Notably, we found no flare signatures in the helium, magnesium, and iron lines, but did find an asymmetrical structure in both sodium lines, and a delayed signal in the manganese lines. Both signals are in line with expectations based on prior work \citep{Doyle1992, Rutten11}. As a result, a modified asymmetrical sodium index was proposed. It is unknown what could cause the signal in the manganese line, given the current consensus that these lines are not pumped. 

\textit{Radial velocity and equivalent width measures.}
In terms of RV measurements, we found that flares of this magnitude can cause RV offsets despite not being visible in TSI. However, these offsets are no larger than a few tens of centimeters per second. This can be explained by the RV signal mainly coming from the wings of photospheric lines (where the flux derivative is sharper), which are formed in a deeper atmospheric layer that is less affected by the flare. 

We find the same for the effects of the flare on the EW measurements, in the sense that it does not seem energetic enough to leave an imprint on the EWs and consequently on the abundances. 

\subsection{Solar context}
In this section, we give a detailed description of the environment that gave rise to the flares, their evolution, and other related signatures resulting from the flares. 

\textit{X-ray flux and (extreme) ultraviolet light curves.}
The disk-integrated emission from AIA full-disk observations is presented as light curves in Fig.~\ref{Fig:aia}. 
Each AIA channel observes a different set of ions formed at different temperatures and thus at different formation heights. This gives rise to another manifestation of the Neupert effect, where the impulsive peak of the flare in each of the channels will occur later or even earlier than the peak of the GOES X-ray flux. This is generally interpreted as being the result of the downward precipitation of energy from the heated corona to lower layers via non-thermal electrons that lose their energy by colliding with ions in the dense lower atmosphere. This creates hard X-rays and drives chromospheric evaporation, that is, the conversion of cool chromospheric material into hot upflowing plasma, which increases the density in the corona and leads to an increase in soft X-ray emission \citep[e.g.,][]{Qiu_2021}. Thus, the ascending part of the SXR curve (before its maximum) is expected to follow approximately the cumulative integral of the impulsive HXR signal \citep{Neupert1968}. \citet{Woods_2011} shows that the 304~\AA\ emission rises early along with the 171~\AA\ emission in the absence of coronal dimming. Coronal dimming can be seen between the two flares but not after (see the 171~\AA\ signal in Fig.~\ref{Fig:aia} between 09:40 and 12:00~UT). The 94~\AA\ emission, on the other hand, tends to peak a few minutes after the peak of the GOES X-ray flux and can have a secondary peak minutes later due to post-flare loop reconnection \citep[e.g.,][]{Mitra_2018}.

\textit{Location and evolution.}
Active region NOAA~12673 appeared as a single symmetrical sunspot on 29~August 2017, which evolved into a complex $\delta$-spot during its disk passage due to a pile-up of trailing spots with opposite polarities. This resulted in a spot with a curved polarity inversion line (PIL) that coincided with a light bridge \citep{Yang2017, Verma2018, Wang2018, Romano2018, Bamba2020}. The region exhibited strong photospheric motions, which when combined with the consistent flux emergence made this active uniquely capable of producing flares \citep{Joshi2022}. The region became one of the most productive active regions of Solar Cycle~24, unleashing in total 40 C-, 20 M-, and four X-class flares \citep{Sun2017, Romano2018, 2021Vissers}. 

The X2.2 flare was the result of a strong shearing flow along the PIL between the dominant positive-polarity region 
and the parasitic negative-polarity umbra encroaching on its east side. 
This caused the intrusion and eventual penetration of the negative-polarity region into the positive polarity in its northern section \citep{Yang2017, Hou2018, Inoue_2018, Romano2018, Wang2018, Verma2018, Inoue_2021, Zou2019, Zou_2020}. 
Figure~\ref{fig:X22}a shows the total signed circular polarisation signal (that is, Stokes-V) in the \CaIR line, which thus displays the chromospheric imprint of the line-of-sight (LOS) magnetic field structure in the early stages of the X2.2 flare at 09:04:30~UT. The flare ribbons over the parasitic, negative-polarity region appear white and impinge on the positive-polarity region to the north of the flare (with flare ribbons that appear black). 
Thus, we interpret that the intrusion of the negative-polarity patch has triggered reconnection that causes brightenings in a number of strongly curved flare ribbons around this impinging polarity as well as in a ribbon to the lower right of this in Fig.~\ref{fig:X22}a. A co-temporal \Halpha COCOPLOT is shown in Fig.~\ref{fig:X22}b, where several bright flare ribbons (with broad emission, hence appearing white) have been formed around the edges of the umbrae (black) in this complex $\delta$-spot (with some overlying cool material).

\citet{Liu2018} and \cite{2019Price} found that the magnetic helicity increased significantly, which is consistent with our observations of the rotation of the impinging negative-polarity umbra (compare Fig.~\ref{fig:X22}a,c and the top row of Fig.~\ref{Fig:bsams}). This rotation occurs co-temporally with the expansion of the flare ribbon areas, and a small white-light flare signature (red arrow in Fig.~\ref{fig:X22}d) indicates deeper energetic action, visible in the \CaK continuum channel and which sweeps around the umbra in a clockwise direction. The ribbon to the south also expands in area (red arrows in Fig.~\ref{fig:X22}e,f, which shows the \Halpha and \CaK COCOPLOTs at 09:11:52~UT), and a small cool arcade rises (magenta loops, see blue arrow).
Where the umbra is impinging, the footpoint of a filament is located. This filament of cool material runs from north to south, along the main PIL of the subsequent X9.3 flare (green arrow, magenta in Fig.~\ref{fig:X22}e,f). 
The flux rope eventually erupted, triggering the X9.3 flare at 11:53~UT \citep{Jiang_2018, Chakraborty21}. 
At times between the two large flares, several smaller brightening can be seen which do not produce any measurable signal in the GOES curve. This suggests that they are either weak flares or below or heating caused by the movement or shuffling of the loops \citep{Skan2023}.

The ribbons of the X9.3 flare were observed from an early stage in the \Halpha 6563~\AA, \CaIR, and \CaK 3933~\AA\ lines and the 4000~\AA\ continuum by the SST (Fig.~\ref{fig:X93}). Flare ribbons form along a great length of the PIL between the two regions. The ribbons start out close to the PIL (Fig.~\ref{fig:X93}a) producing strong central enhancement in the \CaIR line profiles, which are thus bright green in the COCOPLOT. 
Along the PIL, between the ribbons, we also see the previously formed flare filament in dark purple (red arrows in Fig.~\ref{fig:X93}a). 
The LOS magnetic structure before the flare is shown in the HMI magnetogram in  Fig.~\ref{fig:X93}b. Early in the X9.3 flare, a white light signature \citep{Kretzschmar2011} was detected in the \CaK continuum. 

The ribbons expand in area both in the north-south direction as well as outward from the PIL in this stage of the flare, as the reconnection continues and the footpoints of the reconnection field loops expand outward in line with the standard solar flare model. 
Fig.~\ref{fig:X93}c,d display the flare 5~min later at around 12:06~UT in \CaIR and \Halpha respectively. The flare ribbons have continued expanding outward from the PIL, which produces a signature in the HARPS-N data (see Fig.~\ref{Fig:activityintx9}). This will also expand the footpoint areas of chromospheric evaporation, that is, hot upflows from the lower atmosphere that fills the loop with dense hot material and leads to UV and X-ray emission. 

Figure~\ref{fig:X93}e shows a series of `fringes' emitted to the west and south from the western flare ribbon (red arrow). These are the fronts of the sunquake \citep{Zharkov_2020, 2020Zharkova} reported in \citet{Quinn19}. This sunquake coincides in time and propagation direction with the start of an expansion of the flare ribbons and thus potentially corresponds to magnetic reconnection in a new region. This expansion of the ribbon is seen in Fig.~\ref{fig:X93}f at 12:45:21~UT. The western ribbon extends to cover the umbra of both sunspots, one can also observe cool coronal upflows and downflows superimposed over the ribbon along very widely arched or open field lines. This rapid expansion and covering of the umbra cause a peak in brightness in both the GOES and AIA channels (see Fig.~\ref{Fig:aia}). 


\begin{figure}[t]
	
	\centering
	\includegraphics[width=8cm]{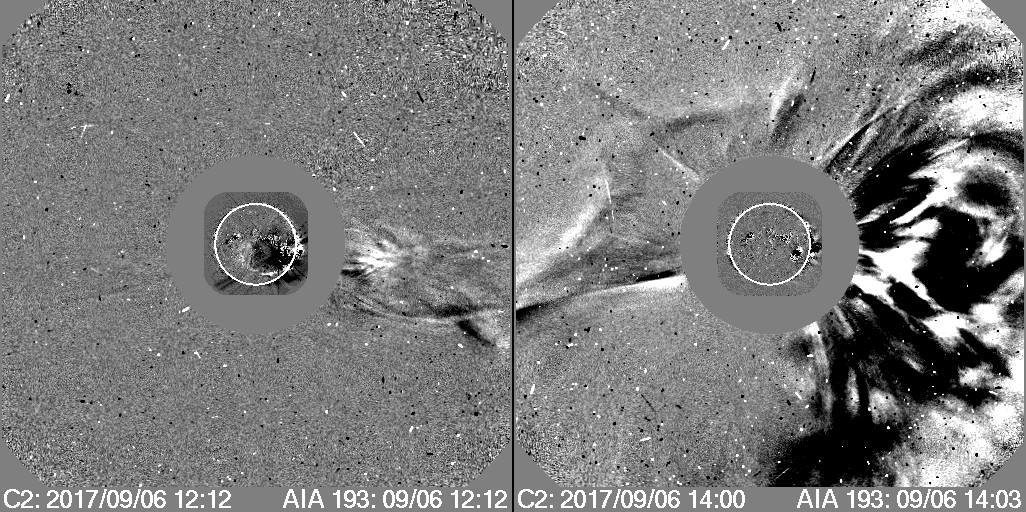}
	\caption{CME related to both flares as seen roughly two hours after their occurrence: (Pseudo)CME corresponding to the X2.2 flare (\textit{left}) and halo-CME corresponding to the X9.3 flare (\textit{right}).}
	\label{Fig:lasco}

\end{figure}

\textit{Corresponding coronal mass ejections.}
The three-di\-men\-sional propagation, and thus the Doppler signal of CMEs in interplanetary space is determined by their launch direction from the Sun, their initial width, and their velocity in interplanetary space. 
Therefore, it is possible to have large CMEs without a corresponding Doppler signal if the launch direction is perpendicular to the Sun-Earth axis, or is launched on the back side of the Sun \citep{Tassev2017}. 
Strong (X-class) flares are eruptive in 90\% of the observed cases. In the remaining 10\%, a flare is confined and produces a so-called pseudo-CME \citep{Vourlidas2010, Vourlidas13}, which is a type of CME that peaks in size below 7~$R_{\odot}$, rather than above 10~$R_{\odot}$, as is usual. A flare's `eruptiveness' is related to its location with respect to the flux-weighed magnetic center of the source active region, with eruptive flares taking place further from the center than confined ones. This is likely due to the overlying arcade field, which dampens the upward motion, thus preventing the flare from becoming eruptive \citep{Wang2007, Thalmann2015}. 

In the case of our flares, the first resulted in a pseudo-CME with an estimated speed of 391~km~s$^{-1}$, while the latter was accompanied by a halo-CME (see Fig.~\ref{Fig:lasco}) with a speed of 1910~km~s$^{-1}$ \citep{scolini2020}.


\textit{Earth response.}
A CME will affect the Earth only when it is launched roughly in our direction, or its expansion is sufficiently large that it appears as a halo CME. The geomagnetic storms resulting from flares and associated CMEs produced by active region NOAA~12673 have been extensively studied in the context of space weather, especially in relation to their effect on Earth. The halo CME launched on September~6 was caused by the X9.3 flare and interacted with two CMEs in interplanetary space. These were previously launched in September~4 from the same active region and reached Earth on September~8, causing a geomagnetic storm \citep{scolini2020}. When a CME hits Earth, it compresses the magnetosphere, magnetically reconnects with the magnetosphere, and injects particles into the van Allen radiation belts. A portion of the injected particles precipitates into the upper atmosphere and causes aurora. However, more importantly, the injected particles increase the strength of Earth's ring current, which again induces a magnetic field that can be measured at the surface. The Dst index gives the strength of the induced magnetic field as determined by four magnetic observatories located near the equator, and is the standard measure for the strength of geomagnetic storms. Our complex CME event resulted in a (provisional) Dst index of \SI{-122}{nT}, classifying the geomagnetic storm on the weaker side of intense geomagnetic storms. 

Besides the geomagnetic storm, the X-ray emission of both flares caused a measurable enhancement in the ionospheric (150\,--\,300~km) electron density and temperature \citep{Yamauchi2018, Yasyukevich2018, Mendoza2019, dePaula2022}. This increase in density absorbs radio waves, including those used by aviation and emergency services \citep{Redmon_2018}, as well as GPS services \citep{Desai2020}. The larger flare caused a 1.5-hour ionosonde blackout in several sun-lit regions, while the smaller one caused a blackout for half the time \citep{Berdermann_2018, Zhang_2019,Fagundes2020,Amaechi_2021}. However, the radio emission from the flare itself did not affect any services in a measurable way \citep{Zhou2018}. 

The X8.3 flare that happened several days after these two had much less of an effect on the Earth due to its proximity to the solar limb \citep{Shagimuratov2020, Chakraborty21}, suggesting that the majority of the emitted light moves radially outward. \citet{Bagiya2018} confirms a type of center-to-limb (CLV) relation between the electron content measured in the sunlit ionosphere and the flare location, which matters more than the strength. \citet{Qian2019} shows the same for the extreme ultraviolet (EUV), but notes that the soft X-ray enhancement is essentially not affected by the location of a flare on the solar disk, which is in line with how GOES measures flare intensities. \cite{Lyakhov_2018} concludes that purely the GOES flare classification without location information of the flare on the disk is not enough to accurately predict the ionosphere response, which can cause the radio wave impact to be underestimated. On top of that, simulated responses on Mars and Venus show differences in the ionospheric response compared to Earth, which means that the results for our planet are likely universal \citep{Yan2022}. Flares, including the X9.3 flare, have also been correlated to earthquakes on Earth \citep{Novikov2020,Marchitelli2020}, although the process is still debated.

These effects on Earth are, however, very mild compared to the potential effects that large stellar flares can have on exoplanets. For example, a strong enough flare can alter, or even erode exoplanet atmospheres, and even cause loss of potential oceans on the surface \citep[and references within]{Ilin2021}. However, the same work showed that the strongest flares tend to appear at high latitudes on the stellar disk, which means that the effects will likely be less intense than an equivalent flare near the disk center.

\subsection{Solar-stellar context}
While much can be learned from the HARPS-N contrast profiles, the prior discussion makes it clear that there are limitations to observing flares in one dimension versus two dimensions. 

For example, both flares have a very similar signature in the S-index, RV profile, and even for chromospheric evaporation, despite being almost an order of magnitude apart in energy. With the duration of flares being the only potential discriminator between their strength. More importantly, it was impossible to distinguish between the eruptive flare that resulted in a halo-CME and the confined flare. This could also explain why it has so-far been rare to detect CME signatures in stellar flare data, as the geometry of the problem might be a strong factor in whether or not the CME is measured. 

For example, the flare measured by \citet{Namekata2021} likely occurred near the disk center, as its contrast profile is similar to the one shown in Fig.~2 of \citet{Otsu_2022}. This shape is explained in \citet{Otsu_2022} as being the result of an interplay between the ejected material going up, and later down, along the LOS. A statistical study on solar flares in a Sun-as-a-star setting could shed more light on the fraction of flares that seem to have an accompanying CME versus the actual fraction. 

In addition, the resolved images showed that the response in the \CaHK lines, and thus the S-index scale primarily with the area of the bright flare ribbons rather than the peak of the flare itself. This response is highly non-linear in relation to flare strength, with potentially even smaller flares producing a similar excess in the spectrum. The \Halpha line and index do not suffer from this problem, and therefore are a much less biased metric.

\section{Conclusions}
\label{sec:conclusion}

In this study, we investigated the properties of an eruptive X9.3 flare and its confined X2.2 predecessor in a one-dimensional disk-integrated setting and compared these findings to high-resolution resolved observations of the same event. While these flares are some of the strongest solar flares ever recorded, they are simultaneously amongst the weakest stellar flares ever detected with energies of roughly 2.2$\times$10$^{31}$ and 9.3$\times$10$^{31}$~erg, respectively. This unique comparative analysis between the HARPS-N disk-integrated spectra and high-resolution observations revealed several valuable insights which are summarized below.

\subsection{Activity indices}
The strongest signal was found in the \CaHK lines, where the enhancement in both the S-index and the contrast profiles was above 1\%. However, the intensity and width of these metrics were comparable for both flares, despite the almost one order of magnitude difference in energy, as well as the fact that the X2.2 flare was confined and the X9.3 flare was eruptive. In addition, the delayed peak of the S-index seemingly aligns with the moment where the flare ribbon covers the most area on-disk, not the peak of the flare itself. This implies that this index is more sensitive to the area of the bright flare ribbons, rather than the total flare brightness. This combined with the fact that small noise-like peaks in this index seemingly align with weak flares, implies that the S-index is highly responsive to flare activity, but in a non-linear way that biases the results towards higher activity. 

This is not the case for other indices, especially the \Halpha line which seems to be a better metric for flare strength, duration, and activity in general. In addition, this line brightened several minutes before the start of the flare, with a time offset consistent with previous research. A similar behavior was noted in the AIA 193, 304, 1600, and 1700~\AA\ channels, which makes these lines/channels potentially valuable as short-term flare predictors. 


We studied multiple other activity indices, including \ion{He}{I}\,D$_3$, the \ion{Na}{I}\,D lines, \ion{Mg}{I} 5173~\AA, \ion{Fe}{I} 6173~\AA, and \ion{Mn}{I} 4031~\AA. Notably, we found no flare signatures in the helium, magnesium, and iron lines, but did find an asymmetrical structure in both sodium lines, and a delayed signal in the manganese lines. Both signals are in line with expectations based on prior work \citep{Doyle1992, Rutten11}, and as a result, a modified asymmetrical sodium index was proposed. It is unknown what could cause the flare signal to present itself in the manganese line.

\subsection{Radial velocity and equivalent width measurements}
In terms of RV measurements, we found that flares of this size can cause RV offsets, but not larger than a few tens of centimeters per second. 
Furthermore, the timing of the peaks aligns better with that of the S-index, meaning that it is more sensitive to the total area of brightenings induced by the flare, rather than the total flare intensity. However, it is hard to draw any strong conclusions from a signal this weak, and a statistical follow-up is required.

Similar conclusions can be drawn regarding the effects of the flare on the EW measurements. The observed flare events seem to be not energetic enough to see signatures in the EWs measurements and consequently on the results of the stellar abundance analysis similar to what is found in much younger stars. Nevertheless, we cannot exclude that flares might be one of the causes (if there are more than one) of the growing EW with activity. It would be interesting to repeat the same study as the time series of a stronger stellar flare to study its effects on the RVs and EWs.

\subsection{Flare signatures and coronal mass ejections}
All prior metrics that register the X2.2 flare do so in a manner that is nearly identical to the X9.3 flare in all aspects except the duration of the enhancement in the contrast profiles. This is despite the fact that the two flares are almost one order of magnitude apart in energy, as well as one being eruptive and the other confined. 

The signature in the contrast profile time series had the same shape as the limb flare studied by \citet{Otsu_2022} and the signature found by \citet{Namizaki2023}. This, together with the findings of \citet{Otsu_2022} and \citet{Gonzalez2022} shows that the location of the flare impacts the shape of the contrast profiles. If this behavior is indeed universal, then a more systematic study into these patterns may result in additional spatial constraints for stellar flares. The same could also shed more light on the rate of false negatives of flare CMEs, because in our case no difference in chromospheric evaporation was seen between the two flares.

\subsection{Outlook}
The recent increased interest in Sun-as-a-star studies is being fuelled by the insights that they provide to both the solar and stellar research communities. They enable solar physicists to contextualize the behavior of the Sun within the broader stellar regime and provide a resolvable solar context for stellar spectral emissions. While our results were fundamentally enabled by the unique nature of this data set, which contained two very strong and well-understood flares, similar investigations can be done for other flares, or even different types of active regions as long as there is some context for them.

Besides small FOV observations, such as those made by the SST, instruments that observe the full disk can also be used for this type of research (e.g., the Chromospheric Telescope \citep[ChroTel,][]{Kentischer2008, Bethge2011}, the Meudon spectroheliogram \citep{Malherbe(2019), Malherbe(2023)}, SDO/AIA\,\&\,HMI \cite{2012LemenAIA, 2012ScherrerHMI}, SOLIS/ISS \citep{Keller2003, Bertello2011}, SMART/SDDI \citep{UeNo2004, Ichimoto2017}, PoET \citep{Leite2022L}, and CHASE \citep{Li2019, Li2022}).

In addition, new Sun-as-a-star instruments are being developed and installed around the globe on telescopes with high-resolution spectrographs (e.g., PEPSI \citep{Strassmeier2018}, HARPS/HELIOS, NEID \citep{Lin(2022)}, ESPRESSO/POET, and EXPRES \citep{Llama2023}). Once all operational, this will result in a (near) constant coverage of the Sun (if weather permits), and thus greatly increase the overlap with general solar observations. 

\begin{acknowledgements}
AP, MV, and CD acknowledge support by grants from the European Commission’s Horizon 2020 Program under grant agreements 824064 (ESCAPE -- European Science Cluster of Astronomy \& Particle Physics ESFRI Research Infrastructures) and 824135 (SOLARNET -- Integrating High-Resolution Solar Physics).
MD is supported by FWO project G0B4521N.
SJH acknowledges support from the German Science Fund under project number 448336908.
KP and EMAG acknowledge support from the German \textit{Leibniz-Gemeinschaft} under project number P67/2018.
IK is supported by KO~6283/2-1 of the Deutsche Forschungsgemeinschaft (DFG).
X.D has received funding from the European Research Council (ERC) under the European Union’s Horizon 2020 research and innovation programme (grant agreement SCORE $N_{0}$851555) and from the Swiss National Science Foundation under grant SPECTRE $200021_215200$
This work has been carried out within the framework of the NCCR PlanetS supported by the Swiss National Science Foundation under grants $51NF40_182901 and 51NF40_205606$.
We thank Gerry Doyle for the stimulating discussion on flare signals in the manganese lines, and the anonymous referee for their valuable suggestions during the peer-review process. 
The Swedish 1-meter Solar Telescope is operated on the island of La Palma by the Institute for Solar Physics of Stockholm University in the Spanish Observatorio del Roque de los Muchachos of the Instituto de Astrof\'isica de Canarias. The Institute for Solar Physics was supported by a grant for research infrastructures of national importance from the Swedish Research Council (registration number 2017-00625).
We acknowledge the VIRGO Experiment; on the cooperative ESA/NASA Mission SoHO from VIRGO Team through PMOD/WRC, Davos, Switzerland for the availability of the TSI data.
We thank the geomagnetic observatories (Kakioka [JMA], Honolulu and San Juan [USGS], Hermanus [RSA], Alibag [IIG]), NiCT, INTERMAGNET, and many others for their cooperation in making the provisional Dst index available. 
This research has made use of NASA's Astrophysics Data System (ADS) bibliographic services. 
We acknowledge the community efforts devoted to the development of the following open-source packages that were used in this work: numpy (\href{http:\\numpy.org}{numpy.org}), matplotlib (\href{http:\\matplotlib.org}{matplotlib.org}), and astropy (\href{http:\\astropy.org}{astropy.org}).
We extensively used the CRISPEX analysis tool \citep{Gregal12}, the ISPy library \citep{ISPy2021}, SunPy \citep{sunpy}, CRISPy \citep{pietrow19}, and SOAImage DS9 \citep{2003DS9} for data visualization. 
After the hard pandemic times the authors want to thank the initiative for social environment improvement at AIP. A substantial part of this paper was envisioned during the social interactions taking place during the 'Scotch Hour' of the AIP. 
\end{acknowledgements} 

\bibliographystyle{aa}
\bibliography{ref}

\end{document}